

%
%
%
\def\unredoffs{} \def\redoffs{\voffset=-.31truein\hoffset=-.59truein}
\def\speclscape{\special{ps: landscape}}
%
%
%
%
\newbox\leftpage \newdimen\fullhsize \newdimen\hstitle \newdimen\hsbody
\tolerance=1000\hfuzz=2pt
\catcode`\@=11 
\def\bigans{b }
\def\answ{b }

%
\ifx\answ\bigans\message{(This will come out unreduced.}
\magnification=1200\unredoffs\baselineskip=16pt plus 2pt minus 1pt
\hsbody=\hsize \hstitle=\hsize 
\else\message{(This will be reduced.} \let\l@r=L
\magnification=1000\baselineskip=16pt plus 2pt minus 1pt \vsize=7truein
\redoffs \hstitle=8truein\hsbody=4.75truein\fullhsize=10truein\hsize=\hsbody
\output={\ifnum\pageno=0 
  \shipout\vbox{\speclscape{\hsize\fullhsize\makeheadline}
    \hbox to \fullhsize{\hfill\pagebody\hfill}}\advancepageno
  \else
  \almostshipout{\leftline{\vbox{\pagebody\makefootline}}}\advancepageno
  \fi}
\def\almostshipout#1{\if L\l@r \count1=1 \message{[\the\count0.\the\count1]}
      \global\setbox\leftpage=#1 \global\let\l@r=R
 \else \count1=2
  \shipout\vbox{\speclscape{\hsize\fullhsize\makeheadline}
      \hbox to\fullhsize{\box\leftpage\hfil#1}}  \global\let\l@r=L\fi}
\fi
%
\newcount\yearltd\yearltd=\year\advance\yearltd by -1900

\def\Title#1#2{\nopagenumbers\abstractfont\hsize=\hstitle\rightline{#1}%
\vskip 1in\centerline{\titlefont #2}\abstractfont\vskip .5in\pageno=0}
\def\Date#1{\vfill\leftline{#1}\tenpoint\supereject\global\hsize=\hsbody%
\footline={\hss\tenrm\folio\hss}}
%

\def\draftmode{\message{ DRAFTMODE }\def\draftdate{{\rm preliminary draft:
\number\month/\number\day/\number\yearltd\ \ \hourmin}}%
\headline={\hfil\draftdate}\writelabels\baselineskip=20pt plus 2pt minus 2pt
 {\count255=\time\divide\count255 by 60 \xdef\hourmin{\number\count255}
  \multiply\count255 by-60\advance\count255 by\time
  \xdef\hourmin{\hourmin:\ifnum\count255<10 0\fi\the\count255}}}
\def\nolabels{\def\wrlabeL##1{}\def\eqlabeL##1{}\def\reflabeL##1{}}
\def\writelabels{\def\wrlabeL##1{\leavevmode\vadjust{\rlap{\smash%
{\line{{\escapechar=` \hfill\rlap{\sevenrm\hskip.03in\string##1}}}}}}}%
\def\eqlabeL##1{{\escapechar-1\rlap{\sevenrm\hskip.05in\string##1}}}%
\def\reflabeL##1{\noexpand\llap{\noexpand\sevenrm\string\string\string##1}}}
\nolabels
%
\global\newcount\secno \global\secno=0
\global\newcount\meqno \global\meqno=1
\def\newsec#1{\global\advance\secno by1\message{(\the\secno. #1)}
\global\subsecno=0\eqnres@t\noindent{\bf\the\secno. #1}
\writetoca{{\secsym} {#1}}\par\nobreak\medskip\nobreak}
\def\eqnres@t{\xdef\secsym{\the\secno.}\global\meqno=1\bigbreak\bigskip}
\def\sequentialequations{\def\eqnres@t{\bigbreak}}\xdef\secsym{}
\global\newcount\subsecno \global\subsecno=0
\def\subsec#1{\global\advance\subsecno by1\message{(\secsym\the\subsecno. #1)}
\ifnum\lastpenalty>9000\else\bigbreak\fi
\noindent{\it\secsym\the\subsecno. #1}\writetoca{\string\quad
{\secsym\the\subsecno.} {#1}}\par\nobreak\medskip\nobreak}
\def\appendix#1#2{\global\meqno=1\global\subsecno=0\xdef\secsym{\hbox{#1.}}
\bigbreak\bigskip\noindent{\bf Appendix #1. #2}\message{(#1. #2)}
\writetoca{Appendix {#1.} {#2}}\par\nobreak\medskip\nobreak}
%
%
\def\eqnn#1{\xdef #1{(\secsym\the\meqno)}\writedef{#1\leftbracket#1}%
\global\advance\meqno by1\wrlabeL#1}
\def\eqna#1{\xdef #1##1{\hbox{$(\secsym\the\meqno##1)$}}
\writedef{#1\numbersign1\leftbracket#1{\numbersign1}}%
\global\advance\meqno by1\wrlabeL{#1$\{\}$}}
\def\eqn#1#2{\xdef #1{(\secsym\the\meqno)}\writedef{#1\leftbracket#1}%
\global\advance\meqno by1$$#2\eqno#1\eqlabeL#1$$}
%
\newskip\footskip\footskip14pt plus 1pt minus 1pt 
\def\footnotefont{\ninepoint}\def\f@t#1{\footnotefont #1\@foot}
\def\f@@t{\baselineskip\footskip\bgroup\footnotefont\aftergroup\@foot\let\next}
\setbox\strutbox=\hbox{\vrule height9.5pt depth4.5pt width0pt}
\global\newcount\ftno \global\ftno=0
\def\foot{\global\advance\ftno by1\footnote{$^{\the\ftno}$}}
%
\newwrite\ftfile
\def\footend{\def\foot{\global\advance\ftno by1\chardef\wfile=\ftfile
$^{\the\ftno}$\ifnum\ftno=1\immediate\openout\ftfile=foots.tmp\fi%
\immediate\write\ftfile{\noexpand\smallskip%
\noexpand\item{f\the\ftno:\ }\pctsign}\findarg}%
\def\footatend{\vfill\eject\immediate\closeout\ftfile{\parindent=20pt
\centerline{\bf Footnotes}\nobreak\bigskip\input foots.tmp }}}
\def\footatend{}
%
%
\global\newcount\refno \global\refno=1
\newwrite\rfile
\def\ref{[\the\refno]\nref}
\def\nref#1{\xdef#1{[\the\refno]}\writedef{#1\leftbracket#1}%
\ifnum\refno=1\immediate\openout\rfile=refs.tmp\fi
\global\advance\refno by1\chardef\wfile=\rfile\immediate
\write\rfile{\noexpand\item{#1\ }\reflabeL{#1\hskip.31in}\pctsign}\findarg}
\def\findarg#1#{\begingroup\obeylines\newlinechar=`\^^M\pass@rg}
{\obeylines\gdef\pass@rg#1{\writ@line\relax #1^^M\hbox{}^^M}%
\gdef\writ@line#1^^M{\expandafter\toks0\expandafter{\striprel@x #1}%
\edef\next{\the\toks0}\ifx\next\em@rk\let\next=\endgroup\else\ifx\next\empty%
\else\immediate\write\wfile{\the\toks0}\fi\let\next=\writ@line\fi\next\relax}}
\def\striprel@x#1{} \def\em@rk{\hbox{}}
\def\lref{\begingroup\obeylines\lr@f}
\def\lr@f#1#2{\gdef#1{\ref#1{#2}}\endgroup\unskip}

\def\addref#1{\immediate\write\rfile{\noexpand\item{}#1}} 
\def\startrefs#1{\immediate\openout\rfile=refs.tmp\refno=#1}
\def\xref{\expandafter\xr@f}\def\xr@f[#1]{#1}
\def\refs#1{\count255=1[\r@fs #1{\hbox{}}]}
\def\r@fs#1{\ifx\und@fined#1\message{reflabel \string#1 is undefined.}%
\nref#1{need to supply reference \string#1.}\fi%
\vphantom{\hphantom{#1}}\edef\next{#1}\ifx\next\em@rk\def\next{}%
\else\ifx\next#1\ifodd\count255\relax\xref#1\count255=0\fi%
\else#1\count255=1\fi\let\next=\r@fs\fi\next}
%

%
\newwrite\ffile\global\newcount\figno \global\figno=1
\def\fig{fig.~\the\figno\nfig}
\def\nfig#1{\xdef#1{fig.~\the\figno}%
\writedef{#1\leftbracket fig.\noexpand~\the\figno}%
\ifnum\figno=1\immediate\openout\ffile=figs.tmp\fi\chardef\wfile=\ffile%
\immediate\write\ffile{\noexpand\medskip\noexpand\item{Fig.\ \the\figno. }
\reflabeL{#1\hskip.55in}\pctsign}\global\advance\figno by1\findarg}
\def\xfig{\expandafter\xf@g}\def\xf@g fig.\penalty\@M\ {}
\def\figs#1{figs.~\f@gs #1{\hbox{}}}
\def\f@gs#1{\edef\next{#1}\ifx\next\em@rk\def\next{}\else
\ifx\next#1\xfig #1\else#1\fi\let\next=\f@gs\fi\next}
\newwrite\lfile
{\escapechar-1\xdef\pctsign{\string\%}\xdef\leftbracket{\string\{}
\xdef\rightbracket{\string\}}\xdef\numbersign{\string\#}}

\def\writestop{\def\writestoppt{\immediate\write\lfile{\string\pageno%
\the\pageno\string\startrefs\leftbracket\the\refno\rightbracket%
\string\def\string\secsym\leftbracket\secsym\rightbracket%
\string\secno\the\secno\string\meqno\the\meqno}\immediate\closeout\lfile}}
\def\writestoppt{}\def\writedef#1{}
\def\seclab#1{\xdef #1{\the\secno}\writedef{#1\leftbracket#1}\wrlabeL{#1=#1}}
\def\subseclab#1{\xdef #1{\secsym\the\subsecno}%
\writedef{#1\leftbracket#1}\wrlabeL{#1=#1}}
\newwrite\tfile \def\writetoca#1{}
\def\leaderfill{\leaders\hbox to 1em{\hss.\hss}\hfill}
\def\writetoc{\immediate\openout\tfile=toc.tmp
   \def\writetoca##1{{\edef\next{\write\tfile{\noindent ##1
   \string\leaderfill {\noexpand\number\pageno} \par}}\next}}}

%
\catcode`\@=12 
%
\edef\tfontsize{\ifx\answ\bigans scaled\magstep3\else scaled\magstep4\fi}
\font\titlerm=cmr10 \tfontsize \font\titlerms=cmr7 \tfontsize
\font\titlermss=cmr5 \tfontsize \font\titlei=cmmi10 \tfontsize
\font\titleis=cmmi7 \tfontsize \font\titleiss=cmmi5 \tfontsize
\font\titlesy=cmsy10 \tfontsize \font\titlesys=cmsy7 \tfontsize
\font\titlesyss=cmsy5 \tfontsize \font\titleit=cmti10 \tfontsize
\skewchar\titlei='177 \skewchar\titleis='177 \skewchar\titleiss='177
\skewchar\titlesy='60 \skewchar\titlesys='60 \skewchar\titlesyss='60
\def\titlefont{\def\rm{\fam0\titlerm}
\textfont0=\titlerm \scriptfont0=\titlerms \scriptscriptfont0=\titlermss
\textfont1=\titlei \scriptfont1=\titleis \scriptscriptfont1=\titleiss
\textfont2=\titlesy \scriptfont2=\titlesys \scriptscriptfont2=\titlesyss
\textfont\itfam=\titleit \def\it{\fam\itfam\titleit}\rm}
 \ifx\answ\bigans\else scaled\magstep1\fi
\ifx\answ\bigans\def\abstractfont{\tenpoint}\else
\font\abssl=cmsl10 scaled \magstep1
\font\absrm=cmr10 scaled\magstep1 \font\absrms=cmr7 scaled\magstep1
\font\absrmss=cmr5 scaled\magstep1 \font\absi=cmmi10 scaled\magstep1
\font\absis=cmmi7 scaled\magstep1 \font\absiss=cmmi5 scaled\magstep1
\font\abssy=cmsy10 scaled\magstep1 \font\abssys=cmsy7 scaled\magstep1
\font\abssyss=cmsy5 scaled\magstep1 \font\absbf=cmbx10 scaled\magstep1
\skewchar\absi='177 \skewchar\absis='177 \skewchar\absiss='177
\skewchar\abssy='60 \skewchar\abssys='60 \skewchar\abssyss='60
\def\abstractfont{\def\rm{\fam0\absrm}
\textfont0=\absrm \scriptfont0=\absrms \scriptscriptfont0=\absrmss
\textfont1=\absi \scriptfont1=\absis \scriptscriptfont1=\absiss
\textfont2=\abssy \scriptfont2=\abssys \scriptscriptfont2=\abssyss
\textfont\itfam=\bigit \def\it{\fam\itfam\bigit}\def\footnotefont{\tenpoint}%
\textfont\slfam=\abssl \def\sl{\fam\slfam\abssl}%
\textfont\bffam=\absbf \def\bf{\fam\bffam\absbf}\rm}\fi
\def\tenpoint{\def\rm{\fam0\tenrm}
\textfont0=\tenrm \scriptfont0=\sevenrm \scriptscriptfont0=\fiverm
\textfont1=\teni  \scriptfont1=\seveni  \scriptscriptfont1=\fivei
\textfont2=\tensy \scriptfont2=\sevensy \scriptscriptfont2=\fivesy
\textfont\itfam=\tenit \def\it{\fam\itfam\tenit}\def\footnotefont{\ninepoint}%
\textfont\bffam=\tenbf \def\bf{\fam\bffam\tenbf}\def\sl{\fam\slfam\tensl}\rm}
\font\ninerm=cmr9 \font\sixrm=cmr6 \font\ninei=cmmi9 \font\sixi=cmmi6
\font\ninesy=cmsy9 \font\sixsy=cmsy6 \font\ninebf=cmbx9
\font\nineit=cmti9 \font\ninesl=cmsl9 \skewchar\ninei='177
\skewchar\sixi='177 \skewchar\ninesy='60 \skewchar\sixsy='60
\def\ninepoint{\def\rm{\fam0\ninerm}
\textfont0=\ninerm \scriptfont0=\sixrm \scriptscriptfont0=\fiverm
\textfont1=\ninei \scriptfont1=\sixi \scriptscriptfont1=\fivei
\textfont2=\ninesy \scriptfont2=\sixsy \scriptscriptfont2=\fivesy
\textfont\itfam=\ninei \def\it{\fam\itfam\nineit}\def\sl{\fam\slfam\ninesl}%
\textfont\bffam=\ninebf \def\bf{\fam\bffam\ninebf}\rm}
%
%

\hyphenation{anom-aly anom-alies coun-ter-term coun-ter-terms}
\def\inv{^{\raise.15ex\hbox{${\scriptscriptstyle -}$}\kern-.05em 1}}

\def\Dsl{\,\raise.15ex\hbox{/}\mkern-13.5mu D} 
\def\dsl{\raise.15ex\hbox{/}\kern-.57em\partial}

 \def\Tr{{\rm Tr}}
\font\bigit=cmti10 scaled \magstep1
\def\lspace{\ifx\answ\bigans{}\else\qquad\fi}
\def\lbspace{\ifx\answ\bigans{}\else\hskip-.2in\fi} 
\def\boxeqn#1{\vcenter{\vbox{\hrule\hbox{\vrule\kern3pt\vbox{\kern3pt
    \hbox{${\displaystyle #1}$}\kern3pt}\kern3pt\vrule}\hrule}}}
\def\mbox#1#2{\vcenter{\hrule \hbox{\vrule height#2in
        \kern#1in \vrule} \hrule}}  
%

\def\darr#1{\raise1.5ex\hbox{$\leftrightarrow$}\mkern-16.5mu #1}

\def\roughly#1{\raise.3ex\hbox{$#1$\kern-.75em\lower1ex\hbox{$\sim$}}}

\let\includefigures=\iftrue
\let\useblackboard=\iftrue
\newfam\black

\includefigures
\message{If you do not have epsf.tex (to include figures),}
\message{change the option at the top of the tex file.}
\input epsf
\def\figin{\epsfcheck\figin}\def\figins{\epsfcheck\figins}
\def\epsfcheck{\ifx\epsfbox\UnDeFiNeD
\message{(NO epsf.tex, FIGURES WILL BE IGNORED)}
\gdef\figin##1{\vskip2in}\gdef\figins##1{\hskip.5in}
\else\message{(FIGURES WILL BE INCLUDED)}%
\gdef\figin##1{##1}\gdef\figins##1{##1}\fi}
\def\DefWarn#1{}
\def\figinsert{\goodbreak\midinsert}
\def\ifig#1#2#3{\DefWarn#1\xdef#1{fig.~\the\figno}
\writedef{#1\leftbracket fig.\noexpand~\the\figno}%
\figinsert\figin{\centerline{#3}}\medskip\centerline{\vbox{
\baselineskip12pt\advance\hsize by -1truein
\noindent\footnotefont{\bf Fig.~\the\figno:} #2}}
\endinsert\global\advance\figno by1}
\else
\def\ifig#1#2#3{\xdef#1{fig.~\the\figno}
\writedef{#1\leftbracket fig.\noexpand~\the\figno}%
\global\advance\figno by1} \fi

\def\id{{1 \kern-.28em {\rm l}}}

\def\K3{{\bf K3}}
\def\journal#1&#2(#3){\unskip, \sl #1\ \bf #2 \rm(19#3) }
\def\andjournal#1&#2(#3){\sl #1~\bf #2 \rm (19#3) }

\def\bar{\overline}
\def\hat{\widehat}
\def\ie{{\it i.e.}}
\def\eg{{\it e.g.}}

\def\tilde{\widetilde}

\def\frac#1#2{{#1\over#2}}

\def\inbar{\,\vrule height1.5ex width.4pt depth0pt}
\def\IC{\relax\hbox{$\inbar\kern-.3em{\rm C}$}}
\def\IR{\relax{\rm I\kern-.18em R}}
\def\IP{\relax{\rm I\kern-.18em P}}

%
%

%
\catcode`\@=11
\def\slash#1{\mathord{\mathpalette\c@ncel{#1}}}
\overfullrule=0pt

\def\AA{{\cal A}}

\def\HH{{\cal H}}

\def\KK{{\cal K}}

\def\OO{{\cal O}}

\def\RR{{\cal R}}

\def\ZZ{{\cal Z}}

\def\underrel#1\over#2{\mathrel{\mathop{\kern\z@#1}\limits_{#2}}}

\catcode`\@=12


%

\def\det{{\rm det}}

\def\Tr{{\rm Tr}}

\def \sgn{{\rm sgn}}
\def\det{{\rm det}}


\def\p{{\partial}}

\lref\CalabreseC{
P.~Calabrese and J.~L.~Cardy,
``Entanglement Entropy and Quantum Field Theory,"
J.\ Stat.\ Mech.\ {\bf 0406}, P002 (2004) [arXiv:hep-th/0405152].
}

\lref\HolzheyWE{
  C.~Holzhey, F.~Larsen and F.~Wilczek,
  ``Geometric and renormalized entropy in conformal field theory,''
Nucl.\ Phys.\ B {\bf 424}, 443 (1994).
[hep-th/9403108].
}

\lref\CalabreseCT{
P.~Calabrese and J.~L.~Cardy,
``Entanglement Entropy and Conformal Field Theory,"
J.\ Phys.\ A {\bf 42} 504005 (2009)[arXiv:0905.4013 [cond-mat.stat-mech]].
}

\lref\CalabreseCM{
J.~L.~Cardy and P.~Calabrese,
``Unusual Correction to Scaling in Entanglement Entropy,"
J.\ Stat.\  Mech. {\bf 1004} P04023 (2010)[arXiv:1002.4353 [cond-mat.stat-mech]].
}

\lref\CasiniFH{
  H.~Casini, C.~D.~Fosco and M.~Huerta,
  ``Entanglement and alpha entropies for a massive Dirac field in two dimensions,''
J.\ Stat.\ Mech.\ {\bf 0507}, P00077 (2005).
[arXiv:cond-mat/0505563].
}

\lref\Casinihm{
  H.~Casini and M.~Huerta,
  ``Entanglement and alpha entropies for a massive scalar field in two dimensions,''
J.\ Stat.\ Mech.\ {\bf 0512}, P12012 (2005).
[arXiv:cond-mat/0511014].
}

\lref\RyuBV{
  S.~Ryu and T.~Takayanagi,
  ``Holographic derivation of entanglement entropy from AdS/CFT,''
Phys.\ Rev.\ Lett.\  {\bf 96}, 181602 (2006).
[hep-th/0603001].
}

\lref\RyuEF{
  S.~Ryu and T.~Takayanagi,
  ``Aspects of Holographic Entanglement Entropy,''
JHEP {\bf 0608}, 045 (2006).
[hep-th/0605073].
}

\lref\RosenhausWOA{
  V.~Rosenhaus and M.~Smolkin,
  ``Entanglement Entropy: A Perturbative Calculation,''
[arXiv:1403.3733 [hep-th]].
}

\lref\RosenhausZZA{
  V.~Rosenhaus and M.~Smolkin,
  ``Entanglement Entropy for Relevant and Geometric Perturbations,''
[arXiv:1410.6530 [hep-th]].
}

\lref\DixonQV{
  L.~J.~Dixon, D.~Friedan, E.~J.~Martinec and S.~H.~Shenker,
  ``The Conformal Field Theory of Orbifolds,''
Nucl.\ Phys.\ B {\bf 282}, 13 (1987).
}

\lref\PolchinskiRR{
  J.~Polchinski,
  ``String theory. Vol. 2: Superstring theory and beyond,''
Cambridge, UK: Univ. Pr. (1998) 531 p.
}

\lref\EinhornUZ{
  M.~B.~Einhorn,
  ``Form-Factors and Deep Inelastic Scattering in Two-Dimensional Quantum Chromodynamics,''
Phys.\ Rev.\ D {\bf 14}, 3451 (1976).
}

\lref\CallanPS{
  C.~G.~Callan, Jr., N.~Coote and D.~J.~Gross,
  ``Two-Dimensional Yang-Mills Theory: A Model of Quark Confinement,''
Phys.\ Rev.\ D {\bf 13}, 1649 (1976).
}

\lref\KatzBR{
  E.~Katz and T.~Okui,
  ``The 't Hooft model as a hologram,''
JHEP {\bf 0901}, 013 (2009).
[arXiv:0710.3402 [hep-th]].
}

\lref\tHooftHX{
  G.~'t Hooft,
  ``A Two-Dimensional Model for Mesons,''
Nucl.\ Phys.\ B {\bf 75}, 461 (1974).
}

\lref\CasiniSR{
  H.~Casini and M.~Huerta,
  ``Entanglement entropy in free quantum field theory,''
J.\ Phys.\ A {\bf 42}, 504007 (2009).
[arXiv:0905.2562 [hep-th]].
}

\lref\CasiniCT{
  H.~Casini and M.~Huerta,
  ``A $c$-theorem for entanglement entropy,''
J.\ Phys.\ A {\bf 40}, 7031-7036 (2007).
cond-mat/0610375v2.
}

\lref\AntonyanQY{
  E.~Antonyan, J.~A.~Harvey and D.~Kutasov,
  ``The Gross-Neveu Model from String Theory,''
Nucl.\ Phys.\ B {\bf 776}, 93 (2007).
[hep-th/0608149].
}

\lref\AntonyanVW{
  E.~Antonyan, J.~A.~Harvey, S.~Jensen and D.~Kutasov,
  ``NJL and QCD from string theory,''
[hep-th/0604017].
}

\lref\KlebanovWS{
  I.~R.~Klebanov, D.~Kutasov and A.~Murugan,
  ``Entanglement as a probe of confinement,''
Nucl.\ Phys.\ B {\bf 796}, 274 (2008).
[arXiv:0709.2140 [hep-th]].
}

\lref\LewkowyczMW{
  A.~Lewkowycz,
  ``Holographic Entanglement Entropy and Confinement,''
JHEP {\bf 1205}, 032 (2012).
[arXiv:1204.0588 [hep-th]].
}

\lref\VelytskyRS{
  A.~Velytsky,
  ``Entanglement entropy in d+1 SU(N) gauge theory,''
Phys.\ Rev.\ D {\bf 77}, 085021 (2008).
[arXiv:0801.4111 [hep-th]].
}

\lref\GromovKIA{
  A.~Gromov and R.~A.~Santos,
  ``Entanglement Entropy in 2D Non-abelian Pure Gauge Theory,''
Phys.\ Lett.\ B {\bf 737}, 60 (2014).
[arXiv:1403.5035 [hep-th]].
}

\lref\DonnellyGVA{
  W.~Donnelly,
  ``Entanglement entropy and nonabelian gauge symmetry,''
Class.\ Quant.\ Grav.\  {\bf 31}, no. 21, 214003 (2014).
[arXiv:1406.7304 [hep-th]].
}

\lref\DattaSKA{
  S.~Datta, J.~R.~David, M.~Ferlaino and S.~P.~Kumar,
  ``Higher spin entanglement entropy from CFT,''
JHEP {\bf 1406}, 096 (2014).
[arXiv:1402.0007 [hep-th]].
}

\lref\DattaZPA{
  S.~Datta, J.~R.~David and S.~P.~Kumar,
  ``Conformal perturbation theory and higher spin entanglement entropy on the torus,''
[arXiv:1412.3946 [hep-th]].
}

\lref\HeadrickFK{
  M.~Headrick, A.~Lawrence and M.~Roberts,
  ``Bose-Fermi duality and entanglement entropies,''
J.\ Stat.\ Mech.\  {\bf 1302}, P02022 (2013).
[arXiv:1209.2428 [hep-th]].
}

\lref\DonnellyFUA{
  W.~Donnelly and A.~C.~Wall,
  ``Entanglement entropy of electromagnetic edge modes,''
[arXiv:1412.1895 [hep-th]].
}

\lref\CasiniRBA{
  H.~Casini, M.~Huerta and J.~A.~Rosabal,
  ``Remarks on entanglement entropy for gauge fields,''
Phys.\ Rev.\ D {\bf 89}, 085012 (2014).
[arXiv:1312.1183 [hep-th]].
}

\lref\ZamolodchikovGT{
  A.~B.~Zamolodchikov,
  ``Irreversibility of the Flux of the Renormalization Group in a 2D Field Theory,''
JETP Lett.\  {\bf 43}, 730 (1986), [Pis'ma Zh.\ Eksp.\ Teor.\ Fiz.\  {\bf 43}, 565 (1986)].
}

\lref\FrishmanZZ{
  Y.~Frishman and J.~Sonnenschein,
  ``Non-perturbative field theory: From two-dimensional conformal field theory to QCD in four dimensions,''
Cambridge, UK: Univ. Pr. (2010) 436 p.
}

\lref\HuangPFA{
  K.~W.~Huang,
  ``Central Charge and Entangled Gauge Fields,''
[arXiv:1412.2730 [hep-th]].
}

\lref\KabatJQ{
  D.~N.~Kabat, S.~H.~Shenker and M.~J.~Strassler,
  ``Black hole entropy in the O(N) model,''
Phys.\ Rev.\ D {\bf 52}, 7027 (1995).
[hep-th/9506182].
}

 \Title{} {\vbox{\centerline{On entanglement entropy in 't~Hooft model}
}}
\bigskip

\centerline{\it   
Mikhail Goykhman
}
\bigskip
\smallskip
\centerline{Enrico Fermi Institute, University of Chicago}
\centerline{5620 S. Ellis Av., Chicago, IL 60637, USA}

\vglue .3cm

\bigskip

\let\includefigures=\iftrue
\bigskip
\noindent
We use the replica trick approach to calculate, in the large-$N$ limit,
the entanglement $c$-function in two-dimensional 
quantum chromodynamics with the gauge group $U(N)$ and quark in the fundamental
representation ('t~Hooft model). We show explicitly that the result is equal to a sum (with
certain numerical coefficients) of
entanglement $c$-functions of free massive scalars, which represent mesons
in spectrum of the 't~Hooft model. Interestingly, each meson contributes
as a point-like particle. We provide an explanation for why this is the case.
\bigskip

\Date{January 2015}

\newsec{Introduction}

The subject of entanglement entropy in quantum field theory received a considerable
attention over the past several years. A general physical setup is the following. One takes
a system in the vacuum state $|0\rangle$, and specifies a sub-region $\AA$ in the coordinate space.
This way the whole system is split
into sub-region $\AA$ and its complement $\bar\AA$.\foot{Below in this section we discuss subtlety of such splitting in
systems with gauge interaction.}
The state of the subsystem $\AA$ is described
by a density matrix $\rho$, obtained by tracing out the degrees of freedom of the subsystem $\bar\AA$,
\eqn\rhodef{\rho={\rm Tr}_{\bar\AA}\,|0\rangle\langle 0|\,.}
The problem is to find corresponding entanglement entropy
\eqn\EEdefI{S=-{\rm Tr}\,\rho\,\log\rho\,,}
where trace is taken over the states of $\AA$.

One way to calculate entanglement entropy \EEdefI\ is a replica trick \refs{\HolzheyWE,\CalabreseC,\CalabreseCT}.
The simplest system, which this method can be applied to, is
a single interval in $2d$ conformal field theory.
If $L$ is a length of the interval, and $c$ is a central charge of the CFT,
then entanglement entropy is given by \refs{\HolzheyWE,\CalabreseC,\CalabreseCT}
\eqn\CFTsiEE{S_{CFT}=\frac{c}{3}\log\,\frac{L}{\epsilon}\,,}
where $\epsilon$ is a UV cutoff. The entanglement $c$-function for
single interval in $2d$ field theory is defined as
\eqn\cdef{c_E=L\,\frac{dS}{dL}\,,}
and due to \CFTsiEE, for CFT it is equal to
\eqn\CFTcfun{c_E=\frac{c}{3}\,.}

Moving away from the conformal regime quickly complicates calculation
of the entanglement entropy.
The question which one
can ask is what is the entanglement entropy of a single interval in the
theory of free $2d$ scalar, or fermion, with a mass $m$. In both cases entanglement entropy is expressed in terms
of solution of the Painlev\'e V equation. This solution is known numerically for general $mL$, and
is known analytically only in asymptotic UV and IR
regions \refs{\CasiniFH,\Casinihm,\CasiniSR}.

It becomes more challenging to find entanglement entropy
in interacting quantum field theory. One of the computational tools has been
suggested by Ryu and Takayanagi \refs{\RyuBV,\RyuEF}, and it gives a prescription to find
entanglement entropy in strongly interacting quantum field theory, which has
a known gravity dual. The other, purely field theoretical, method
involves perturbative computation of entanglement entropy, see, \eg, \refs{\RosenhausWOA,\RosenhausZZA}
and references therein, for recent developments.

In this paper we use the replica trick approach to calculate entanglement
$c$-function in two-dimensional quantum chromodynamics
with a single-flavored quark in the fundamental representation of the gauge group $U(N)$.
This system is known as 't~Hooft model \tHooftHX.
It is solvable in the large-$N$ limit.
The spectrum of the 't~Hooft model is given by a tower of mesons, with masses
of sufficiently heavy mesons approximated by the formula
\eqn\intR{m_k^2\simeq  \pi^2 \lambda k\,,}
where $\lambda$ is the 't~Hooft coupling \refs{\tHooftHX,\CallanPS}, see \FrishmanZZ\ for a review.
The mesons are non-interacting in the large-$N$ limit.

Entanglement entropy in pure $U(N)$ gauge theory in two dimensions has been calculated and
discussed, \eg, in
\refs{\VelytskyRS,\GromovKIA,\DonnellyGVA}.
There is a subtlety in defining entanglement entropy for systems with gauge interaction.
One way to see the problem is to look at gauge-invariant operators of a finite size.
The question is, how to treat the gauge-invariant operators which are partly located
in $\AA$, and partly in $\bar\AA$. 
Following \refs{\VelytskyRS,\GromovKIA,\DonnellyGVA} we assume that entanglement entropy is well-defined in $2d$
field theory with gauge interaction, and replica trick calculation is applicable, and gives a sensible answer.
(See also \CasiniRBA\ and references therein for discussion of entanglement entropy
in gauge theories.)

Each meson in the spectrum of the 't~Hooft model can be viewed as a string of
a given tension. Length of such a string increases as mass of the meson grows.
The subtlety in defining entanglement entropy
arises, \eg, when we look at string which has one end inside the interval $\AA$,
and the other end outside of the interval.
At the same time, due to having a finite length,
one would expect that mesons contribute to the entanglement entropy as finite-size
objects.
However, our replica trick calculation shows that each meson contributes to the entanglement $c$-function as
a point-like particle. We explain that this is a consequence of a consistent definition of \DonnellyGVA\ of
entanglement entropy in gauge theories.

In two dimensions entanglement $c$-function \cdef\ is a monotonically
decreasing function of the (dimensionless)
length of the interval \CasiniCT. Therefore it can be compared with the Zamolodchikov $c$-function.
The latter is defined via the two-point function of the stress-energy tensor \refs{\ZamolodchikovGT,\PolchinskiRR}. Schematically, it is given by
\eqn\Zamcfun{C(|z|)\simeq z^4\langle T(z)T(0)\rangle\,.}
In 't~Hooft model we have qualitative behavior of Zamolodchikov $c$-function given by
\foot{See App. B of \KatzBR\ for results on two-point functions of general bi-fermionic
operators in the 't~Hooft model.}
\eqn\expecfun{C(\sqrt{\lambda}|z|)=N\, F(\sqrt{\lambda}|z|)\,,}
where $F$ is a smooth function of order one, monotonically decreasing between one
in the ultra-violet and zero in the infra-red.
At a fixed point Zamolodchikov $c$-function is equal to central charge of the corresponding CFT.
Similarly, entanglement $c$-function of CFT is proportional to the central
charge \CFTcfun.

Due to these similarities between entanglement and Zamolodchikov $c$-functions,
we expect the entanglement $c$-function for the 't~Hooft model to behave in a way, similar to \expecfun.
However, our calculation demonstrates
a different behavior. In agreement with \CasiniCT\ we have obtained a monotonically
decreasing entanglement $c$-function.
We have derived that enanaglement $c$-functions is $c_E(\sqrt{\lambda}L)=\OO(1)$
for $\lambda L^2\gg 1/N$, unlike the Zamolodchikov $c$-function \expecfun.
When $\lambda L^2=1/N$, we obtained that the entanglement $c$-function approaches
the value of $\OO(N)$.
In this regime interaction of quarks is parametrically suppressed, and perturbative
calculation of entanglement entropy should be done.\foot{See, \eg, \refs{\DattaSKA,\DattaZPA} for examples of perturbative replica trick calculation of
entanglement entropy in $2d$ quantum field theory.}
We leave this for future work.

As a warm-up, we start by analyzing a simpler theory with four-fermionic
interaction, the Thirring model. This model is well known to be equivalent
to a free massless scalar. This can be immediately demonstrated by bosonization
of fermions. Therefore the single-interval entanglement entropy in Thirring model
is given by the CFT formula \CFTsiEE\ with central charge $c=1$.
We choose to postpone bosonization, and apply the replica trick approach
to calculation of entanglement entropy in the Thirring model.
We explicitly show that CFT result \CFTsiEE\ for free complex-valued fermion ($c=1$)
is not changed by Thirring four-fermionic interaction.

Before proceeding to the 't~Hooft model, we study a simpler gauge theory
in two dimensions, Schwinger model of a fermion interacting with a $U(1)$ gauge field.
Gauge field in two dimensions is non-dynamical and can be explicitly
integrated over. This generates a non-local four-fermionic interaction.
Unlike the case of Thirring model, in this case we have a relevant interaction,
and it affects the result for the entanglement entropy.

It is well known that Schwinger
model is equivalent to free massive scalar.
We assume that single-interval entanglement entropy
of Schwinger model maps to single-interval entanglement entropy of
massive scalar field theory. It is not obvious that such an assumption
is correct, because of non-local nature of duality between Schwinger model
and massive scalar field \DonnellyGVA.
Similar subtlety in relation between entanglement entropy in
bosonic and fermionic pictures can appear even in CFT.
It appears already in the case of multiple-interval entanglement entropy,
which is not defined by simple expression analogous to \CFTsiEE, in CFTs
of free boson and free fermion \HeadrickFK. Careful treatment shows
that the results for multiple-interval entanglement entropy do indeed agree
for the theories, related by bosonization \HeadrickFK.

We match the expression
for the entanglement entropy in Schwinger model
with the result of \Casinihm\ for massive scalar field. This gives an expression, which will be useful in calculation
of entanglement entropy in the 't~Hooft model.
We show that calculation of entanglement entropy in the 't~Hooft model 
and the Schwinger model goes through similar steps.
By explicit calculation
we show that single-interval entanglement
$c$-function in the 't~Hooft model (in the large-$N$ limit) is given by a sum (with certain numerical coefficients) of entanglement $c$-functions of free scalars
with masses $m_k$ in the spectrum of the 't~Hooft model.


The rest of this paper is organized as follows. In Section 2 we summarize the
formulae for entanglement entropy, and outline the general calculation
procedure which we will follow. We start with the CFT of free fermions, and then proceed to
deforming the CFT by an arbitrary operator.
We discuss in general how such an operator contributes to the single-interval entanglement entropy.
In Section 3 we calculate entanglement
entropy in the Thirring model, testing the formulae derived in Section 2. In Section 4 we discuss entanglement entropy in the Schwinger model,
and derive expression, which will be useful for calculation in the 't~Hooft model. In Section 5 we derive expression for
entanglement entropy in the 't~Hooft model. We discuss our results in Section 6, where we provide
an argument for why each meson, now matter how heavy it is, contributes to the entanglement
entropy as a point-like particle.
Appendix A is dedicated to derivation of gap equation governing dressed fermionic propagator in the 't~Hooft model.
In Appendix B we review derivation of quark-antiquark scattering amplitude in the 't~Hooft model.

\newsec{Preliminaries}

In Subsection 2.1 we collect the defining formulae for entanglement
entropy which we will need in this paper. In Subsection 2.2 we discuss free massless fermions
on $n$-sheeted Riemann surface.
In Subsection 2.3 we perturb the CFT of free fermions by operator $\OO$ and discuss
how it contributes to the entanglement entropy.

\subsec{Entanglement entropy}

Consider $2d$ system in a vacuum state. Suppose the subsystem $\AA$
is a spatial interval of length $L$. Denote $\rho$ to be density matrix
obtained by tracing out all degrees of freedom outside of the $\AA$.
One can compute the trace ${\rm Tr}\,\rho^n$ for positive integer values of $n$.
This quantity enters definition of the R\'enyi entropy
\eqn\Rendef{S_n=\frac{1}{1-n}\log\,\Tr\, \rho^n\,.}
Then, following the replica trick, one analytically continues the expression \Rendef\ to any real value of $n$,
and calculates entanglement entropy using the formula
\eqn\EntR{S=-\Tr\,\rho\log\rho=-\frac{\p}{\p n}\log\,\Tr\,\rho^n\Bigg|_{n=1}=\lim_{n\rightarrow 1}S_n\,.}

Evaluation of entanglement entropy therefore gets reformulated as a problem
of finding the R\'enyi entropy \Rendef. To calculate it, let us put the system on $n$-sheeted Riemann surface $\RR_n$.
Different sheets of $\RR_n$ are glued at the cut $\AA$, and boundary conditions
across the cut are specified.
Then the trace $\Tr\,\rho^n$ can be determined as \refs{\HolzheyWE,\CalabreseC,\CalabreseCT}
\eqn\RhoRn{\Tr\,\rho^n=\frac{Z_n}{Z_1^n}\,,}
where $Z_n$ is partition function on $\RR_n$.

Due to \Rendef, \RhoRn, the R\'enyi entropy can be determined as
\eqn\RenZn{S_n=\frac{1}{1-n}(\log\,Z_n-n\log\,Z_1)\,.}
The entanglement entropy is therefore given by
\eqn\EEe{S=\log\, Z_1-\frac{1}{Z_1}\frac{\p Z_n}{\p n}\Bigg|_{n=1}\,.}
For the future purposes it is useful to express the partition function as
\eqn\ZZndef{Z_n=\frac{1}{L^{2N\Delta_n}}\ZZ(n)\,,}
where $\Delta_1=0$ and
$\frac{\p\Delta_n}{\p n}|_{n=1}=\frac{1}{6}$.
Then entanglement entropy is
\eqn\SZon{S=\frac{N}{3}\log\, L+\log\, Z_1-\frac{1}{Z_1}\frac{\p\ZZ (n)}{\p n}\Bigg|_{n=1}\,.}
To get rid of the logarithm $\log\,Z_1$ in \SZon\ one usually defines entanglement $c$-function
\eqn\cdefon{c_E=L\,\frac{d}{dL}\,S=\frac{N}{3}-L\frac{d}{dL}\,\frac{1}{Z_1}\frac{\p\ZZ (n)}{\p n}\Bigg|_{n=1}\,.}

Suppose we start with conformal field theory with central charge equal to $N$.
Then if we perturb it by some interaction, the entanglement entropy
will be given by \SZon, where the $\frac{N}{3}\log\, L$ term is a CFT input \CalabreseCT, and $\ZZ(n)$ term
originates from non-trivial interaction.

\subsec{Free massless fermions}

Let us consider a free massless fermionic theory on $n$-sheeted Riemann surface $\RR_n$.
We have left-moving (holomorphic) fermion $\Psi=\Psi_1+i\Psi_2$
and right-moving (anti-holomorphic) fermion $\bar \Psi=\bar \Psi_1+i\bar \Psi_2$,
with the free field action
\eqn\Thiract{S=\int _{\RR_n}d^2w\left(\Psi^\star \bar\p\Psi +\bar\Psi^\star \p \bar\Psi\right)\,.}
In our notation bar denotes chirality of the fermion, and star denotes complex conjugation. 
The integral in \Thiract\ is taken over $n$-sheeted Riemann surface $\RR_n$
with the branch points $w_{1,2}$, and the sheets are glued along the cut $[w_1,w_2]$. Without loss of generality, we choose $w_1=0$, $w_2=L$,
and take $L$ to be real-valued.
Instead of considering fermions $(\Psi\,,\bar\Psi)$
on $\RR_n$, consider
$n$ fermions $(\Psi_j \,,\bar\Psi_j)$, $j=1,\dots,n$, living on a complex $z$-plane \CalabreseCT.
The action on the complex plane is
\eqn\ThiractRn{S=\int d^2z\sum_{j=1}^n\left(\Psi_j^\star \bar\p\Psi _j+\bar\Psi^\star_j \p \bar\Psi_j\right)\,,}
and we must insert twist operators $T(0)\tilde T(L)$ into all correlation functions.
Before describing the action of the twist operators,
let us do a conformal transformation of a complex plane,
\eqn\zumap{u=\frac{z-L}{z}\,,}
which maps twist operators to the points $u=0$ and $u=\infty$.
These twist operators lead to the boundary conditions on fermions \CasiniFH\
\eqn\Tact{ \left({\Psi_j\atop \bar\Psi_j}\right)\rightarrow \left({\Psi_{j+1}\atop \bar\Psi_{j+1}}\right)
\,,\quad  \left({\Psi_n\atop \bar\Psi_n}\right)
\rightarrow (-1)^{n+1}\left({\Psi_1\atop \bar\Psi_1}\right)}
when one circles clockwise around $u=0$, and
\eqn\Tactt{ \left({\Psi_j\atop \bar\Psi_j}\right)\rightarrow \left({\Psi_{j-1}\atop \bar\Psi_{j-1}}\right)
\,,\quad  \left({\Psi_1\atop \bar\Psi_1}\right)
\rightarrow (-1)^{n+1}\left({\Psi_n\atop \bar\Psi_n}\right)}
when one circles around $u=\infty$ (\ie, when one circles around $u=0$ counter-clockwise).
The complex $u$-plane can be mapped onto a cylinder with world-sheet coordinates
$(\tau,\sigma)$:
\eqn\tsmap{u=e^{\tau+i\sigma}\,.}
The twist operators are inserted at $\tau=\pm\infty$, at infinite past and infinite future.
The coordinate $\sigma\in [0,2\pi]$ parametrizes circumference of the cylinder.

Let us switch to diagonal basis of twist operators,
by doing a discrete Fourier transform
\eqn\diagbas{\Psi_j=\sum _{s=-\frac{n-1}{2}}^{\frac{n-1}{2}}\frac{1}{\sqrt{n}}
e^{2\pi i j s/n}\psi_s\,,\qquad
\bar\Psi_j=\sum _{s=-\frac{n-1}{2}}^{\frac{n-1}{2}}\frac{1}{\sqrt{n}}
e^{2\pi i j s/n}\bar\psi_s\,.}
In the basis $\psi_s$, $\bar \psi_s$, $s=-\frac{n-1}{2},\dots,\frac{n-1}{2}$ the
boundary conditions are
\eqn\Tactd{ \left({\psi_s\atop \bar\psi_s}\right)\rightarrow e^{2\pi is/n}\left({\psi_s\atop \bar\psi_s}\right)}
when one circles around $u=0$ (goes around circumference of the cylinder), and
\eqn\Tactdt{  \left({\psi_s\atop \bar\psi_s}\right)\rightarrow e^{-2\pi is/n}\left({\psi_s\atop \bar\psi_s}\right)}
when one circles around $u=\infty$ (goes around circumference of the cylinder
in the opposite direction). The action for fermions $(\psi_s\,,\bar\psi_s)$ is
\eqn\CFTfer{S=\int d^2z\sum_{s=-\frac{n-1}{2}}^{\frac{n-1}{2}}\left[\psi^\star _s
\bar\p\psi_s+\bar\psi^\star _s\p\bar\psi_s\right]\,.}

Fermions, satisfying the boundary conditions \Tactd, \Tactdt, have the mode expansion
\eqn\psimode{\eqalign{
\psi_s^\star(u)=\sum_{r\in Z+1/2}\frac{\psi^\star_{(s)\;r+\frac{s}{n}}}{u^{r+\frac{s}{n}+\frac{1}{2}}}\,,\quad
\quad 
\psi_s(u)=\sum_{r\in Z+1/2}\frac{\psi_{(s)\;r-\frac{s}{n}}}{u^{r-\frac{s}{n}+\frac{1}{2}}}\,,\cr
\bar \psi_s^\star(\bar u)=\sum_{r\in Z+1/2}\frac{\bar\psi^\star_{(s)\;r-\frac{s}{n}}}{\bar u^{r-\frac{s}{n}+\frac{1}{2}}}\,,\quad
\quad 
\bar \psi_s(\bar u)=\sum_{r\in Z+1/2}\frac{\bar \psi_{(s)\;r+\frac{s}{n}}}{\bar u^{r+\frac{s}{n}+\frac{1}{2}}}
}}
where the amplitudes of expansion satisfy anti-commutation relations \refs{\DixonQV,\PolchinskiRR}
\eqn\psiacr{\eqalign{
\{\psi^\star_{(s_1)\;r_1+s_1/n},\,\psi_{(s_2)\;r_2-s_2/n}\}=\delta_{s_1,s_2}\delta_{r_1,-r_2}\,,\cr
\{\bar\psi^\star_{(s_1)\;r_1-s_1/n},\,\bar\psi_{(s_2)\;r_2+s_2/n}\}=\delta_{s_1,s_2}\delta_{r_1,-r_2}
}}
and the vacuum is defined as \foot{In our case $s/n$ is never equal to $1/2$,
so we do not have to worry about zero modes: $\psi_{(s)\;0}$ never comes up.}
\eqn\vacdef{\eqalign{\psi ^\star_{(s)\;r+s/n}|0\rangle =0\,,\quad \psi_{(s)\;r+s/n}|0\rangle =0\,,\quad r>0\,,\cr
\bar\psi ^\star_{(s)\;r+s/n}|0\rangle =0\,,\quad \bar\psi_{(s)\;r+s/n}|0\rangle =0\,,\quad r>0\,,}
}
for all $s$.
Using equations \psimode, \psiacr, \vacdef\ we compute the correlation functions
\eqn\psiscorf{\langle \psi^\star_{s_1}(u_1)\psi_{s_2}(u_2)\rangle =\delta_{s_1s_2}
\left(\frac{u_2}{u_1}\right)^\frac{s_1}{n}\frac{1}{u_1-u_2}\,.}
For the anti-holomorphic sector we obtain
\eqn\psiscorfahs{\langle \bar\psi^\star_{s_1}(\bar u_1)\bar \psi_{s_2}(\bar u_2)\rangle =\delta_{s_1s_2}
\left(\frac{\bar u_1}{\bar u_2}\right)^\frac{s_1}{n}\frac{1}{\bar u_1-\bar u_2}\,.}

We can also arrive at the correlation functions \psiscorf, \psiscorfahs\ by computing
the limit of four-point functions of fermions and twist operators
\eqn\sigmafpf{\eqalign{&\lim_{R\rightarrow\infty}\,R^{2\Delta_s}\left\langle \sigma_s^\star(0)\psi_s^\star(u_1)\psi_s(u_2)\sigma_s(R)\right\rangle\cr
&=\frac{1}{u_1-u_2}\lim_{R\rightarrow\infty}\, \left(\frac{u_2}{u_1-R}\frac{u_1-R}{u_1}\right)^{\frac{s}{n}}=
\frac{1}{u_1-u_2}\left(\frac{u_2}{u_1}\right)^\frac{s}{n}\,.}}
In \sigmafpf\ the $\Delta_s=s^2/(2n^2)$ is dimension of twist operator
$\sigma_s(u)=e^{i\frac{s}{n}\phi_s(u)}$. We bosonized the
fermions, $\psi_s(u)=e^{i\phi_s(u)}$,
where $\langle \phi_s(u_1)\phi_s(u_2)\rangle =-\log\,(u_1-u_2)$.
Similarly one can derive correlation function for anti-holomorphic fermions \psiscorfahs.

For the future purposes we also are going to need correlation function for $\Psi_j$
fermions. Due to \diagbas\ and \psiscorf\ we obtain
\eqn\Psicor{\langle \Psi_{j_1}^\star(u_1)\Psi_{j_2}(u_2)\rangle=
(-1)^{j_1+j_2}\frac{1}{n}\frac{e^{\frac{\pi i j_1}{n}}u_1^{\frac{1}{2n}-\frac{1}{2}}
e^{\frac{\pi i j_2}{n}}u_2^{\frac{1}{2n}-\frac{1}{2}}}{e^{\frac{2\pi i j _1}{n}}u_1^{\frac{1}{n}}-
e^{\frac{2\pi i j _2}{n}}u_2^{\frac{1}{n}}}\,.}
Similarly, for anti-holomorphic fermions we obtain
\eqn\Psibarcor{\langle \bar\Psi_{j_1}^\star(\bar u_1)\bar\Psi_{j_2}(\bar u_2)\rangle=
(-1)^{j_1+j_2}\frac{1}{n}\frac{e^{-\frac{\pi i j_1}{n}}\bar u_1^{\frac{1}{2n}-\frac{1}{2}}
e^{-\frac{\pi i j_2}{n}}\bar u_2^{\frac{1}{2n}-\frac{1}{2}}}{e^{-\frac{2\pi i j _1}{n}}\bar u_1^{\frac{1}{n}}-
e^{-\frac{2\pi i j _2}{n}}\bar u_2^{\frac{1}{n}}}\,.}

\subsec{Interacting fermions}

Suppose CFT of free fermions is deformed by the interaction
\eqn\intgen{S_{int}=g\int d^2z \, \OO(z,\bar z)\,,}
where $\OO$ is a composite operator of $\Psi$, $\bar\Psi$, and $g$ is corresponding coupling constant.
Suppose $\OO$ in \intgen\ is operator
of dimensions $(\Delta,\bar\Delta)$. Then conformal transformation of $\OO$ on complex
plane, $w=f(z)$, is given by
\eqn\Oplane{\left(\frac{\p f(z)}{\p z}\right)^{\Delta}
\left(\frac{\bar \p \bar f(\bar z)}{\bar \p \bar z}\right)^{\bar \Delta}\OO(f(z),\bar f(\bar z))=\OO(z,\bar z)\,.}

We can study the model \intgen\ on Riemann surface $\RR_n$
by considering
$n$ copies of the original model on complex $z$-plane,
and specifying boundary conditions for going around singular points $z=0,L$. Let us map the singular
points to $u=0,\infty$ by transformation \zumap.
The corresponding action is then
\eqn\intgenRn{S_{int}=g\int d^2u\,J(u,\bar u)\sum_{j=1}^n \OO_j(u,\bar u)\,,}
and the boundary conditions are
\eqn\OObc{\OO_j(e^{2\pi i}u,e^{-2\pi i}\bar u)= \OO_{j+1}(u,\bar u)\,,\qquad
\OO_n(e^{2\pi i}u,e^{-2\pi i}\bar u)= \OO_{1}(u,\bar u)\,.}
In \intgenRn\ we have introduced the $J(u,\bar u)$, which is a product of Jacobian of coordinate change \zumap\ and factors coming from conformal transformation \Oplane\ of operators $\OO_j$,
\eqn\Jfor{J(u,\bar u)=\frac{L^{2-\Delta-\bar\Delta}(1-u)^{2\Delta}(1-\bar u)^{2\bar \Delta}}{|1-u|^4}\,.}

Notice that boundary conditions \OObc\ are just conditions
for going between sheets of Riemann surface of the $n$th root function $u^{1/n}$. Therefore the sum
over $j$ in \intgenRn\ actually stands for summing over $n$
sheets of $u^{1/n}$. This means that we can do the coordinate transformation $u=v^n$,
mapping $n$ sheets of $u^{1/n}$ onto complex plane with $v$ coordinate.
\foot{Mapping from $\RR_n$ to complex plane is frequently used in calculations
of entanglement entropy, see, \eg, \refs{\CalabreseCT,\CalabreseCM}.}
The way this transformation acts on $\OO$ is described by equation analogous to \Oplane,
\eqn\ORnp{\left(nv^{n-1}\right)^{\Delta}\left(n\bar v^{n-1}\right)^{\bar\Delta}\,\sum _j \OO_j(v^n,\bar v^n)=\OO(v,\bar v)\,.}
Now we can perform $u=v^n$ transformation in the action \intgenRn. Due to \ORnp\ we obtain
\eqn\Sintvv{S_{int}=g\,n^{2-\Delta-\bar\Delta}\int d^2v\,J(v^n,\bar v^n)\,\left(v^{1-\Delta}\,
\bar v^{1-\bar \Delta}\right)^{n-1}\OO(v,\bar v)\,.}

Up to this moment the discussion has been general and did not rely on the fact that $\OO_j$ operators
are composed of $\Psi_j$, $\bar\Psi_j$ fermions. After we do the change of variables $u=v^n$
we obtain operator $\OO(v,\bar v)$ in \Sintvv, composed of single-valued on a plane fermions $\Psi(v)$,
$\bar\Psi(\bar v)$.
The free-field correlation functions of these fermions are
\eqn\Psivcor{\langle \Psi^\star (v_1)\Psi(v_2)\rangle=\frac{1}{v_1-v_2}\,,\qquad
\langle \bar\Psi^\star (\bar v_1)\bar \Psi(\bar v_2)\rangle=\frac{1}{\bar v_1-\bar v_2}\,.}

This conclusion can also be reached in a more explicit way.
First, one writes down perturbative expansion of the partition function
\eqn\Zgen{Z=\int[D\Psi]\, e^{-S_{int}}\,,}
with $S_{int}$ given by \intgenRn. Each term in the perturbative expansion can
be evaluated, by using correlation functions of fermions
\Psicor, \Psibarcor. One notices that it is possible to do the transformation
$u=v^n$ for each given order of the perturbative expansion. Indeed, for example
in holomorphic sector every $u_m^{1/n}$
is multiplied by phase factor $e^{2\pi i j_m/n}$, and $j_m$ is summed over from $1$ to $n$,
see \Psicor.
Then all the terms can be assembled back into partition function of fermions \Psivcor, deformed by
interaction \Sintvv.

\newsec{Thirring model}

As a warm-up, in this section we consider Thirring model,
described by the action
\eqn\Thir{S_T=\int d^2z\left(\Psi^\star\bar\p\Psi+\bar\Psi^\star\p\bar\Psi +\lambda _T
\Psi^\star\Psi\bar\Psi^\star\bar\Psi\right)\,,}
where $\lambda_T$ is a dimensionless coupling constant.
A well-known fact is that one can immediately bosonize the fermions, $\Psi(z)=e^{i\varphi (z)}$,
$\bar \Psi(\bar z)=e^{i\bar \varphi (\bar z)}$, and notice that the action
\Thir\ describes free massless scalar field $\varphi$. Therefore the entanglement
entropy of a single interval in Thirring model is simply given by \CFTsiEE\
with $c=1$. We prefer to postpone bosonization and
instead treat the four-fermionic term in the action \Thir\ as
a perturbation $\OO$. Then we can test the procedure, described in Subsection 2.3.

Let us put the model \Thir\ on $n$-sheeted Riemann surface $\RR_n$. This can be accomplished by dealing with $n$
copies of fermions, $(\Psi_j,\bar\Psi_j)$, $j=1,\dots,n$ on a complex $z$-plane, with the action
\eqn\Thir{S_T=\int d^2z\sum_{j=1}^n\left(\Psi _j^\star\bar\p\Psi_j+\bar\Psi_j^\star\p\bar\Psi _j+\lambda_T 
\Psi_j^\star\Psi_j\bar\Psi_j^\star\bar\Psi_j\right)\,,}
and twist operators inserted at $z=0,L$.
The interaction in \Thir\  is of the form \intgen\ with $(\Delta,\bar \Delta)=(1,1)$.
In the $u$ coordinates \zumap,
the fermions $(\Psi_j,\bar\Psi_j)$ satisfy boundary conditions 
\Tact, \Tactt.

Due to \Jfor\ we obtain $J(u,\bar u)=1$.
The partition function of Thirring model is therefore given by
\eqn\ZnThir{Z_n^T=\frac{1}{L^{2\Delta_n}}\int [D\Psi]\,
e^{-\lambda _T\int d^2u\sum_{j=1}^n\Psi_j^\star\Psi_j\bar\Psi_j^\star\bar\Psi_j}\,,}
where
\eqn\Deltan{\Delta_n=\frac{n^2-1}{12n}\,.}
We can apply conformal perturbation theory for
computation of \ZnThir.
In the previous section we computed CFT correlation functions \Psicor, \Psibarcor\ for the $(\Psi_j,\bar\Psi_j)$ fermions.
We can either use these correlation functions for explicit calculation, or use
general result derived in Subsection 2.3. Due to \Sintvv\ we obtain simply
\eqn\Zzr{Z_n=\frac{1}{L^{2\Delta_n}}
\left\langle e^{-\lambda_T\int d^2v\bar\p\bar\varphi\p\varphi}\right\rangle\,,}
where we have bosonized the fermions, and
\eqn\varphic{\langle \varphi (v_1)\varphi (v_2)\rangle =-\log\, (v_1-v_2)\,,\qquad
\langle \bar\varphi (\bar v_1)\bar\varphi (\bar v_2)\rangle =-\log\, (\bar v_1-\bar v_2)\,.}
From \Zzr\ we conclude that entanglement entropy of Thirring model is given by CFT formula
\CFTsiEE, with $c=1$, as expected.

\newsec{Schwinger model}

In this section we are going to derive formula for entanglement $c$-function in the Schwinger model ($2d$ QED with massless fermion).
Derivation splits into three parts. First, by using the procedure described
in Subsection 2.3, we derive an expression for
$\ZZ(n)$, defined in \ZZndef, which contains all the non-trivial contribution
from the gauge interaction. The resulting formula contains fermionic four-point
function in Schwinger model, which we compute in Subsection 4.2. We assemble all
the results in Subsection 4.3 to write down expression for entanglement $c$-function.

The purpose of this section is twofold. First, it can be viewed as a warm-up for
't~Hooft model: in this section we study
entanglement entropy in $U(1)$ gauge theory with massless fermion, and
in the next section we generalize it to $U(N)$ gauge theory with massless fundamental fermion.
Second, Schwinger model is equivalent to a free massive $2d$ scalar, for which
entanglement entropy is known (numerically) \Casinihm. It turns out that calculation
of entanglement entropy in Schwinger model is very similar to calculation in
't~Hooft model, as we will show in the next Section. Therefore the result for 't~Hooft
model can be expressed using known result for Schwinger model.

\subsec{Derivation of partition function $\ZZ_m(n)$}

The Schwinger model in the $A=0$ gauge, with the gauge coupling $m$, is described by the action
\eqn\actSchw{S=\int d^2z\left(-\frac{1}{2}(\p\bar A)^2+m\bar A\Psi^\star\Psi+\Psi^\star\bar\p\Psi +\bar\Psi^\star
\p\bar\Psi\right)\,.}
The gauge field is non-dynamical in two dimensions, so we can integrate over it.
This amounts to substitution of
\eqn\Abareq{\bar A(z_1,\bar z_1)=m\int d^2z_2\, G(z_1-z_2)\Psi^\star (z_2)\Psi (z_2)\,,}
into the action \actSchw,
where
\eqn\Gzdef{\p_z^2 G(z_1-z_2)=\delta (z_1-z_2)
\quad\Rightarrow\quad G(z_1-z_2)=\frac{1}{2\pi}\frac{z_1-z_2}{\bar z_1-\bar z_2}\,.}
Therefore we can consider Schwinger model as a CFT of free fermions
perturbed by the interaction of holomorphic fermions
\eqn\Sscgo{S_S=m^2\int d^2z_{1,2}\, G(z_1-z_2)\Psi^\star (z_1)\Psi (z_1)\Psi^\star (z_2)\Psi (z_2)\,.}
It is well known that the Schwinger model is equivalent to a free scalar with mass $m$.
This can easily be seen by bosonizaion the fermions. Then \Sscgo\ becomes just a mass
term for a free scalar. As in the case of Thirring model, let us postpone bosonization
for now.

Let us consider Schwinger model on $n$-sheeted Riemann surface $\RR_n$,
with the sheets glued along the cut of the length $L$. As in the case of Thirring model,
this can be accomplished by considering $n$ copies of fermions $(\Psi_j,\,\bar\Psi_j)$ on a complex
plane, with the interaction given by
\eqn\Sscgo{S_S=m^2\int d^2z_{1,2}\,G(z_1-z_2)\, \sum_{j=1}^n
\Psi _j^\star (z_1)\Psi _j(z_1)\Psi^\star _j(z_2)\Psi _j(z_2)\,.}
By doing change of variables \zumap\ we obtain
\eqn\Sscgo{S_S=m^2L^2\int d^2u_{1,2}\, F(u_1,u_2)\sum_{j=1}^n
\Psi _j^\star (u_1)\Psi _j(u_1)\Psi^\star _j(u_2)\Psi _j(u_2)\,,}
where we have denoted
\eqn\Fuudef{F(u_1,u_2)=\frac{1}{|1-u_1|^2|1-u_2|^2}G(u_1-u_2)\,.}
The action \Sscgo\ can be written in a form, similar to \intgenRn,
\eqn\OOschw{S_S=m^2L^2\int d^2u_{1,2}\, F(u_1,u_2)\sum_{j=1}^n\OO_j(u_1)\OO_j(u_2)\,,}
where operators $\OO_j(u)=\Psi^\star_j(u)\Psi_j(u)$ of dimension $(\Delta,\bar\Delta)=(1,0)$ satisfy boundary conditions \OObc.
Therefore for the model \OOschw\ to be well-defined, the coordinates $u_{1,2}$
should go around singular points $u=0,\infty$ at the same time.
This means that the sum over $j$ allows to make change of variables $u_{1,2}=v_{1,2}^n$.
Similarly to \ORnp\ we obtain
\eqn\bilinuvch{n^2v_1^{n-1}v_2^{n-1}\sum _j\OO_j(v_1^n)\OO_j(v_2^n)=\OO(v_1)\OO(v_2)\,,}
where $\OO(v)=\Psi^\star (v)\Psi(v)$, and correlation function for $\Psi (v)$ is given by \Psivcor.
The action therefore becomes
\eqn\Schwact{S_S=n^2m^2L^2\,\int d^2v_{1,2}\, H(v_1,v_2;n)\Psi^\star (v_1)\Psi(v_1)
\Psi^\star (v_2)\Psi(v_2)\,,}
where we denoted
\eqn\Hvvdef{H(v_1,v_2;n)=\frac{\bar v_1^{n-1}\bar v_2^{n-1}}{|1-v_1^n|^2|1-v_2^n|^2}\frac{v_1^n-v_2^n}{
\bar v_1^n-\bar v_2^n}\,.}
Again, one can also arrive at the same result \Schwact\ by using correlation
functions \Psicor, \Psibarcor\ in explicit  perturbative calculation of the partition function.

The partition function of Schwinger model on $\RR_n$ is given by
\eqn\ZnSchw{Z_n=\frac{1}{L^{2\Delta_n}}\int [D\Psi]\, e^{-S_S}\,.}
In \ZnSchw\ we have path integral in CFT of free fermion $\Psi(v)$
with correlation function \Psivcor, perturbed by interaction \Schwact. Following \ZZndef, let us denote
\eqn\Zofns{\ZZ_m(n)=\int [D\Psi]\, e^{-S_S}\,.}
Therefore $\ZZ(1)=Z_1^S$ is just a partition function of free massive $2d$ scalar on a complex plane.
For the purpose of calculation of entanglement entropy \SZon, we are interested in
\eqn\pZZpn{\eqalign{\frac{\p\, \ZZ_m(n)}{\p n}\Bigg|_{n=1}=-m^2L^2\int d^2v_{1,2}\,
H_0(v_1,v_2)\,Z_1^S\,\left\langle\Psi^\star (v_1)\Psi(v_1)
\Psi^\star (v_2)\Psi(v_2)\, e^{-S_S|_{n=1}}\right\rangle\,,}}
where we have denoted
\eqn\Fdef{H_0(v_1,v_2)=2H(v_1,v_2;1)+\frac{\p H(v_1,v_2;n)}{\p n}\Bigg|_{n=1}\,.}
Now we need to compute fermionic four-point function, which enters \pZZpn.

\subsec{Calculation of quark four-point correlation function}

To compute correlation function in \pZZpn, let us bosonize the fermions,
\eqn\pZZpnt{\eqalign{\frac{1}{Z_1^S}\frac{\p\, \ZZ_m(n)}{\p n}\Bigg|_{n=1}=m^2L^2\int d^2v_{1,2}\,
H_0(v_1,v_2)\left\langle \p_{1}\Phi (v_1,\bar v_1)\p_2\Phi (v_2,\bar v_2)\, e^{-S_\Phi}\right\rangle\,,}}
where due to \Schwact, \Hvvdef\ we have
\eqn\Sphi{S_\Phi=-m^2L^2\int d^2v_{1,2}\,\frac{1}{|1-v_1|^2|1-v_2|^2}\frac{v_1-v_2}{\bar v_1-\bar v_2}\;
\p_1\Phi (v_1,\bar v_1)\p_2\Phi (v_2,\bar v_2)\,.}
We can easily get rid of the extra factors in the integral \Sphi\ by doing transformation,
inverse to \zumap:
\eqn\vytr{\frac{y_{1,2}-1}{y_{1,2}}=v_{1,2}\,,}
where $y_{1,2}$ are new dimensionless coordinates on complex plane. This gives
\eqn\Sphiy{S_\Phi=-m^2L^2\int d^2y_{1,2}\;\frac{y_1-y_2}{\bar y_1-\bar y_2}\;
\p_{y_1}\Phi \p_{y_2}\Phi=m^2L^2\int d^2 y\, \Phi (y,\bar y)^2\,,}
where we have also used \Gzdef. Applying the same transformation in \pZZpnt\ gives
\eqn\pZZpntt{\frac{1}{Z_1^S}\frac{\p\, \ZZ(n)}{\p n}\Bigg|_{n=1}=m^2L^2\int d^2y_{1,2}\,
H_1(y_1,y_2)\p_{y_1}\p_{y_2}\left\langle \Phi (y_1,\bar y_1)\Phi (y_2,\bar y_2)\, e^{-S_\Phi}\right\rangle\,,}
where we have denoted
\eqn\hatFdef{H_1(y_1,y_2)=H_0\left(\frac{y_1-1}{y_1},\frac{y_2-1}{y_2}\right)\,\frac{1}{\bar y_1^2\bar y_2^2}\,.}

Let us perform Wick rotation and go to light-cone coordinates,
$(\bar y, y)=(y^+,y^-)$, $(p,\bar p)=(p_+,p_-)$. Then $p\cdot y\equiv p_+y^++p_-y^-=
p\bar y+\bar p y$. We are working in $A_-=0$ gauge. Therefore the residual gauge transformations
are parametrized by $\alpha=\alpha (y^+)$.
In the light-cone $(y^+,y^-)$ coordinates it becomes manifest why \pZZpn\
is a gauge-invariant object, since fermions as not separated in the $y^+$
direction, $y_1^+=y_2^+=y^+$.

Due to \Sphiy\ we have in \pZZpntt\ correlation function of free massive scalar, with
dimensionless mass $m^2L^2$,
\eqn\msccor{\left\langle \Phi (y^+,y_1^-)\Phi (y^+,y_2^-)\, e^{-S_\Phi}\right\rangle= 
\int d^2q\,e^{iq_-(y_1^--y_2^-)}\frac{1}{q^2-m^2L^2}\,.}
Therefore, from \pZZpntt\ we arrive at
\eqn\pZZpntt{\frac{1}{Z_1^S}\frac{\p\, \ZZ_m(n)}{\p n}\Bigg|_{n=1}=m^2L^2\int dy^-_{1,2}\,dy^+\,H_1(y_1,y_2)\,
\int d^2q\,e^{iq_-(y_1^--y_2^-)}\frac{q_-^2}{q^2-m^2L^2}\,.}
The
integral \pZZpntt\ will also appear in the calculation of entanglement entropy in 't~Hooft model.
This integral requires regularization, and as written,
should be viewed formally. In the next Subsection we derive a useful expression for this integral.

\subsec{Entanglement $c$-function}

Due to \SZon\ to find entanglement entropy we need to evaluate \pZZpntt.
This is a hard exercise and, as mentioned  above, requires
regularization.
Fortunately we know the answer for entanglement entropy $S_m(mL)$
of free scalar field with mass $m$ \refs{\Casinihm,\CasiniSR,\CasiniCT}. Therefore we can use the known value of $S_m(mL)$
to find \SZon\
\eqn\pZmm{\frac{1}{Z_1^S}\frac{\p\, \ZZ_m(n)}{\p n}\Bigg|_{n=1}=\frac{1}{3}\log\,L+\log\, Z_1-S_m(mL)\,.}
Differentiating w.r.t. $L$ we obtain
\eqn\pZcmm{L\frac{\p}{\p L}\left[\frac{1}{Z_1^S}\frac{\p\, \ZZ_m(n)}{\p n}\Bigg|_{n=1}\right]
=\frac{1}{3}-c_m(mL)\,,}
where $c_m(mL)$ is entanglement $c$-function of free scalar with mass $m$.
Equations \pZZpntt,
\pZcmm\ give rise to
\eqn\ScU{L\frac{\p}{\p L}\,\left[m^2L^2\int dy^-_{1,2}\,dy^+\,H_1(y_1,y_2)\,
\int d^2q\,e^{iq_-(y_1^--y_2^-)}\frac{q_-^2}{q^2-m^2L^2}\right]=\frac{1}{3}-c_m(mL)\,.}
Equation \ScU\ is the main result of this section, and it will play a key
role in derivation of the entanglement entropy in 't~Hooft model. When $m=0$ we have $c_0=1/3$,
and therefore regularized value of the expression in the l.h.s. of \ScU\ is equal to zero.


\newsec{'t~Hooft model}

In this section we derive expression for entanglement $c$-function in the 't~Hooft model. As in the case
of Schwinger model, derivation splits into three parts: calculation of
$\ZZ(n)$, defined in \ZZndef; calculation of fermionic four-point
function in the 't~Hooft model; and assembling all the results together, to
get the expression for the entanglement $c$-function.
Calculation of $\ZZ(n)$ is similar to analogous calculation, performed for the Schwinger model.

\subsec{Derivation of partition function $\ZZ(n)$}

The 't~Hooft model in the $A^a=0$ gauge is described by the action \tHooftHX\
\eqn\acthooft{S=\int d^2z\left(-\frac{1}{2}(\p\bar A^a)^2+\Psi^{\alpha\, \star}\bar\p\Psi ^\alpha+\bar\Psi^{\alpha \,\star}
\p\bar\Psi^\alpha +g\bar A^a\Psi^{\alpha\, \star} T^a_{\alpha\beta}\Psi^\beta\right)\,.}
Here $T^a$, $a=1,\dots, N^2$ are generators of the $u(N)$ algebra, and $g$ is coupling constant, with dimension of mass.
Fermions $\Psi^\alpha$, $\bar\Psi^\alpha$, $\alpha =1,\dots, N$ live in fundamental representation of the $U(N)$.
Similarly as in the case of Schwinger model, we can integrate over the gauge field,
which amounts to substitution of
\eqn\AsoltH{\bar A^a(z_1)=g\int d^2z_2 \, G(z_1-z_2)\Psi^{\alpha\,\star} (z_2)T^a_{\alpha\beta}\Psi ^\beta(z_2)}
into the action \acthooft. This gives rise
to free fermionic theory, deformed by the interaction
\eqn\inttH{S_H=g^2\,\int d^2z_{1,2}\, G(z_1-z_2)\Psi^{\alpha\star }(z_1)\Psi^{\beta}(z_1)
\Psi^{\beta \star}(z_2)\Psi^{\alpha}(z_2)\,.}
In \inttH\ we have used
\eqn\SuNT{T^a_{\alpha_1\alpha_2}T^a_{\alpha_3\alpha_4}
=\frac{1}{2}\delta_{\alpha_1\alpha_4}\delta_{\alpha_2\alpha_3}
\,.}
The Green's function for $\p^2$ operator is given by \Gzdef. 

Let us consider 't Hooft model on $n$-sheeted Riemann surface $\RR_n$
with branch points at $z=0,L$. Let us make coordinate change \zumap,
and consider $n$
copies of fermions $(\Psi_j^\alpha,\bar\Psi_j^\alpha)$ on a $u$-plane, satisfying boundary conditions \Tact, \Tactt.
These fermions are described by CFT, deformed by the interaction 
\eqn\SatHdef{S_H=g^2 L^2\int d^2u_{1,2}\, F(u_1,u_2)\sum_j \Psi^{\alpha\star }_j(u_1)\Psi^{\beta}_j(u_1)
\Psi^{\beta \star}_j(u_2)\Psi^{\alpha}_j(u_2)\,,}
where $F(u_1,u_2)$ is defined in \Fuudef.

Let us proceed as in the case of Schwinger model. The action \SatHdef\ can be written as
\eqn\SatHOO{S_H=g^2 L^2\int d^2u_{1,2}\, F(u_1,u_2)\sum_j \OO_j^{\alpha\beta}(u_1)\OO^{\beta\alpha}_j(u_2)\,,}
where $\OO^{\alpha\beta}_j(u)=\Psi^{\alpha\star }_j(u_1)\Psi^{\beta}_j(u_1)$.
For this action to be well-defined the $u_{1,2}$ should go around singular points $u_{1,2}=0,\infty$
at the same time. This means that we are allowed to make a change variables $u_{1,2}=v_{1,2}^n$,
mapping $n$ sheets of $u^{1/n}$ function onto complex $v$-plane.
The transformation law of operators $\OO^{\alpha\beta}$ is
given by
\eqn\OOthtr{n^2v_1^{n-1}v_2^{n-1}\sum _j\OO_j^{\alpha\beta}(v_1^n)
\OO_j^{\beta\alpha}(v_2^n)=\OO^{\alpha\beta}(v_1)\OO^{\beta\alpha}(v_2)\,,}
where $\OO^{\alpha\beta}(v)=\Psi^{\alpha\star}(v)\Psi^\beta(v)$.
Correlation function of holomorphic fermions on $v$-plane is
\eqn\Psiab{\langle \Psi^{\alpha\star} (v_1)\Psi^\beta(v_2)\rangle=\delta^{\alpha\beta}\frac{1}{v_1-v_2}\,.}
In the $v$ coordinates the action \SatHOO\ becomes
\eqn\SatHOOv{S_H= n^2 g^2 L^2\int d^2v_{1,2}\, H(v_1,v_2;n)\Psi^{\alpha \star}(v_1)
\Psi^\beta(v_1)\Psi^{\beta\star}(v_2)\Psi^\alpha (v_2)\,,}
where $H(v_1,v_2;n)$ is defined in \Hvvdef.

The partition function of 't~Hooft model on $\RR_n$ is therefore given by
\eqn\ZntH{Z_n=\frac{1}{L^{2N\Delta_n}}\int [D\Psi]\, e^{-S_H}\,.}
In \ZntH\ we have path integral in CFT of free fermions
with correlation functions \Psiab, perturbed by interaction \SatHOOv.
As in the case of Schwinger model, and following \ZZndef, let us denote
\eqn\zzshcdef{\ZZ(n)=\int [D\Psi]\, e^{-S_H}}
in \ZntH. To compute entanglement entropy \SZon, we have to find
\eqn\pZZpntH{\eqalign{\frac{\p\, \ZZ(n)}{\p n}\Bigg|_{n=1}=-g^2L^2\int d^2v_{1,2}\,
H_0(v_1,v_2)\,Z_1^H\,\left\langle\Psi^{\alpha \star} (v_1)\Psi^\beta(v_1)
\Psi^{\beta \star} (v_2)\Psi^\alpha(v_2)\, e^{-S_H|_{n=1}}\right\rangle\,,}}
where $H_0(v_1,v_2)$ is defined in \Fdef, and $Z_1^H$ is partition function of 't~Hooft model on a plane.
In \pZZpntH\ we have $S_H|_{n=1}$,
which is just four-fermionic interaction in 't~Hooft model on complex plane.
Indeed, doing change of coordinates \vytr, we obtain
\eqn\pZZttH{\eqalign{\frac{1}{Z_1^H}\frac{\p\, \ZZ(n)}{\p n}\Bigg|_{n=1}=-g^2L^2\int d^2y_{1,2}\,
H_1(y_1,y_2)\left\langle\Psi^{\alpha \star} (y_1)\Psi^\beta(y_1)\Psi^{\beta \star} (y_2)\Psi^\alpha(y_2)\, \right\rangle _H\,,}}
where $H_1(y_1,y_2)$ is defined in \hatFdef. Subscript $H$ 
stands for correlation function in the theory with interaction
\eqn\Shdef{S^1_H=g^2L^2\int d^2y_{1,2}\,
\frac{y_1-y_2}{\bar y_1-\bar y_2}\;\Psi^{\alpha \star} (y_1)\Psi^\beta(y_1)
\Psi^{\beta \star} (y_2)\Psi^\alpha(y_2)\,.}
We can bring the gauge field back, restoring the original 't~Hooft
model framework, with fermions interacting locally with the gauge field:
\eqn\Shgf{S^1_H=\int d^2y\left(-\frac{1}{2}(\p\bar A^a)^2+\hat g\, \bar A^a\,\Psi^{\alpha\star }T^a_{\alpha\beta}\Psi^\beta\right)\,,}
where
\eqn\hatg{\hat g=gL}
is dimensionless gauge coupling.

\subsec{Calculation of quark four-point correlation function}

In \pZZttH\ we have four-quark amplitude in 't~Hooft model on a plane,
\eqn\Hcf{\KK(y_1,y_2)=
\left\langle\Psi^{\alpha \star} (y_1)\Psi^\beta(y_1)\Psi^{\beta \star} (y_2)\Psi^\alpha(y_2)\right\rangle _H\,.}
Due to translational symmetry, $\KK (y_1,y_2)=\KK(y_1-y_2)$.
This is gauge-invariant quantity, which has been demonstrated
explicitly in \EinhornUZ. The situation is analogous to
Schwinger model, discussed in the previous Section.
Remember that we are working in the gauge $A^a=0$,
and therefore the residual gauge transformations are parametrized by $\alpha^a$
such that
\eqn\gtr{0=\delta A^a=\p \alpha ^a+ f^{abc}A^b\alpha ^c=\p\alpha^a\,,}
that is, $\alpha ^a=\alpha ^a(\bar z)$.
Therefore the object $\Psi^{\alpha\star}(y_1)\Psi^\alpha (y_2)$ is gauge-invariant simply because both fermions
are not separated in the $\bar y$ direction.
Therefore the four-fermionic amplitude in \pZZttH\ is gauge-invariant. In fact one can consider a simpler
gauge-covariant ``blob" \tHooftHX
\eqn\blob{\delta^{\alpha_1\alpha_2}\,\Phi_r(y_1,y_2)=\langle r|\Psi^{\alpha_1 \star} (y_1)\Psi ^{\alpha_2}(y_2)|0\rangle\,,}
giving an amplitude of creation of meson $|r\rangle$ by quark-anti-quark pair $\Psi^{\alpha\star}(y_1)\Psi^\alpha (y_2)$.

The Fourier transform of amplitude \Hcf\ is given by
\eqn\KKft{\KK(y_1-y_2)=\int d^2p_{1,2,3,4}\,e^{i\left(y_1\cdot (p_1-p_2)+y_2\cdot (p_3-p_4)\right)}\,
\left\langle \hat\Psi^{\alpha \star}(p_1)\hat\Psi^\beta (p_2)
\hat\Psi^{\beta\star}(p_3)\hat\Psi ^\alpha (p_4)\right\rangle _H\,.}
Define vertex $T(p,p';q)$ \CallanPS\ for quark-anti-quark scattering as 
\eqn\drverdef{\eqalign{&\left\langle \hat\Psi^{\alpha_1 \star}(p_1)\hat\Psi^{\alpha_2} (p_2)
\hat\Psi^{\alpha_3\star}(p_3)\hat\Psi ^{\alpha_4} (p_4)\right\rangle _H=\delta \left( p_1-p_2+p_3-p_4\right)\, P(p_1)P(p_2)P(p_3)P(p_4)\cr
&\times\left[\,\delta^{\alpha_1\alpha_4}\delta^{\alpha_2\alpha_3}\,T(p_1,p_2;p_1-p_4)-
\delta^{\alpha_1\alpha_2}\delta^{\alpha_3\alpha_4}\,T(p_1,p_4;p_1-p_2)\,\right]
\,,}}
where $P(p)$ is the full quark propagator \refs{\tHooftHX,\CallanPS}.
Let us use \drverdef\ in \KKft, obtaining
\eqn\KKtft{\KK(y_1-y_2)=\KK_1(y_1-y_2)+\KK_2(y_1-y_2)\,,}
where
\eqn\KKftot{\eqalign{\KK_1(y_1-y_2)&=N^2\int d^2p_{1,2,4}\,e^{i(p_1-p_2)\cdot (y_1-y_2)}\,
T(p_1,p_2;p_1-p_4)\cr &\times P(p_1)P(p_2)P(-p_1+p_2+p_4)P(p_4)\,,}}
\eqn\KKft{\eqalign{\KK_2(y_1-y_2)&=N\int d^2p_{1,2,4}\,e^{i(p_1-p_2)\cdot (y_1-y_2)}\,
T(p_1,p_4;p_1-p_2)\cr &\times P(p_1)P(p_2)P(-p_1+p_2+p_4)P(p_4)\,,}}
The $N^2$ in $\KK_1$ appeared after we summed over
$\alpha_1=\alpha_4=\alpha$ and $\alpha_2=\alpha_3=\beta$,
and $N$ in $\KK_2$ appeared after we summed over $\alpha=\beta$.

Let us denote integrated momenta in \KKftot\ as $p_1=p$, $p_2=p'$,
$p_4=p-q$, and define
\eqn\hatRdef{\hat K_1(q)=\int d^2p\,d^2p'\,e^{i(p-p')\,\cdot (y_1-y_2)}\,T(p,p';q)P(p-q)P(p)P(p'-q)P(p')\,.}
Therefore
\eqn\Kyyo{\KK_1(y_1-y_2)=N^2\,\int d^2q\,\hat K_1(q)\,.}

Let us denote integrated momenta in \KKft\ and $p_1=p$, $p_2=p-q$,
$p_4=p'$, and define
\eqn\hatRdeyf{\hat K_2(q)=\int d^2p\,d^2p'\,T(p,p';q)P(p-q)P(p)P(p'-q)P(p')\,.}
Therefore
\eqn\Kyy{\KK_2(y_1-y_2)=N\,\int d^2q\,e^{iq\cdot (y_1-y_2)}\,\hat K_2(q)\,.}
The $\KK_2$ is sub-leading in the large-$N$ limit, and we will not consider it.
In what follows we will only be interested
in $\KK_1$.

Let us perform Wick rotation and switch to light-cone coordinates, 
similarly to what we have done in Subsection 4.2, for Schwinger model. 
Denote $\hat g$ to be gauge coupling constant. The 't~Hooft coupling
constant is $\hat\lambda=\hat gN^2$. The extra hat over the couplings
is introduced because at the end we will substitute $\hat g=gL$,
where $g$ is the original gauge coupling constant, and $L$
is the length of interval, see \hatg\ and preceding discussion.


Let us introduce the IR cutoff $\lambda_{IR}$. Then full quark propagator,
in the large-$N$ limit, is given by
\tHooftHX\
\foot{Equation for full quark propagator $P(p)$ is derived in Appendix A. Quark propagator is two by two matrix. In $A_-=0$ gauge we are only considering entry of this matrix
which corresponds to two-point function of left-moving quarks, since right-moving quarks are non-dynamical.}
\eqn\PLambda{P(p)=\frac{i}{2p_+-\frac{\hat \lambda}{\pi}
\frac{\sgn (p_-)}{\lambda_{IR}}-\frac{M^2-i\epsilon}{p_-}}\,,}
where
\eqn\Mdrdef{M^2=m_q^2-\hat \lambda/\pi\,,}
and $\hat m_q=m_qL$ is dimensionless bare quark mass. At the very end of calculation
of gauge-invariant observables one sends $\lambda_{IR}\rightarrow 0$.

It is known that $T(p,p';q)$, in the large-$N$ limit, satisfies equation \CallanPS\ (see Appendix B, where derivation of this equation
is reviewed)
\eqn\BSTeq{T(p,p';q)=\frac{i\hat g^2}{(p_--p_-')^2}+
\frac{i\hat \lambda}{\pi^2}\int dk_-\,\frac{1}{(k_--p_-)^2}\int dk_+\,P(k)P(k-q)T(k,p';q)\,.}
Denote
\eqn\xxpdef{x=\frac{p_-}{q_-}\,,\qquad x'=\frac{p_-'}{q_-}\,,}
which stand for the fraction of total momentum $q_-$, carried by one of in-coming quarks $p_-$, and one
of the out-going quarks $p_-'$.
The solution to equation \BSTeq\ is \CallanPS\
\eqn\CCGsol{T(p_-,p_-';q)=\frac{\hat \lambda}{N(p_--p_-')^2}-\frac{\hat \lambda^2}{\pi Nq_-^2}
\sum_r\frac{1}{q^2-\hat m_r^2+i\epsilon}\int _0^1dy\int _0^1dy'\frac{\phi _r^\star (y')\phi_r(y)}{(y'-x')^2(y-x)^2}\,.}
Integrals over $y$ and $y'$ in \CCGsol\ are assumed to be taken with the cutoff: holes of size $\lambda_{IR}/q_-$
are drilled around $y=x$ and $y'=x'$.
In \CCGsol\ we denoted $\phi _r(y)$ to be set of eigenfunctions of 't~Hooft equation
\eqn\tHeq{\frac{\pi M^2}{\hat \lambda y(1-y)}\phi_r(y)-\hat P\int _0^1 dy'\frac{\phi _r(y')}{(y'-y)^2}=
\frac{\hat m_r^2}{\hat\lambda}\phi _r(y)\,,}
describing meson of mass $\hat m_r=m_rL$. Here $\hat P$ stands for principal value.

Eigenfunctions $\phi_r(x)$ of 't~Hooft equation \tHeq\
are non-vanishing when $x\in [0,\, 1]$ and satisfy completeness and orthonormality conditions \refs{\tHooftHX, \CallanPS}
\eqn\tHeqsol{\eqalign{\sum_r\,\phi_r(x)\phi_r(x')&=\delta(x-x')\,,\cr
\int _0^1dx\,\phi_r(x)\phi^\star _{r'}(x)&=\delta_{rr'}\,.
}}
For large $r$ (since $N$ does not appear in \tHeq, large $r$ is {\it not} $\OO(N^p)$) we have \refs{\tHooftHX, \CallanPS}
\eqn\philr{\phi_r(x)\simeq \sqrt{2}\sin\,(\pi rx)\,,\qquad \hat m_r^2\simeq \pi^2 \hat\lambda\,r\,.}

Let us re-write \CCGsol\ as \KatzBR\
\eqn\Tsolbs{T(p_-,p_-';q)=\frac{\hat\lambda}{N(p_--p_-')^2}+
\frac{4\hat\lambda^2}{\pi N\lambda_{IR}^2}
\sum_r\frac{1}{q^2-\hat m_r^2+i\epsilon}\chi _r(x,q_-)\chi_r^\star (x',q_-)\,,}
where we have defined
\eqn\chin{\chi_r(x,q_-)=\frac{\lambda_{IR}}{2q_-}\int _0^1dy\frac{\phi_r(y)}{(y-x)^2}\,.}
The reason for such reformulation is that
one can see that presence of $\lambda_{IR}$ in the r.h.s. of \chin\ makes
the $\chi_r$ finite in $\lambda_{IR}\rightarrow 0$ limit \KatzBR. Indeed, we can split the integral over $y$
into principal value part, which is finite when $\lambda_{IR}\rightarrow 0$, and infinite part, coming from divergence near $y=x$:
\eqn\Prvspl{\int _0^1dy\frac{\phi_r(y)}{(y-x)^2}=\hat P \int _0^1dy
\frac{\phi_r(y)}{(y-x)^2}+\frac{2}{\lambda_{IR}/q_-}\phi _r(x)\,.}
We conclude that
\eqn\psinphin{\chi_r(x,q_-)=\phi _r(x)+\OO(\lambda _{IR})\,.}

Now let us discuss $\KK_1(y_1-y_2)$, see \hatRdef, \Kyyo. We have
\eqn\hatRdef{\eqalign{\hat\KK_1(q)=\int dp_-\,dp_-'\,e^{i(p_--p_-')\cdot (y_1^--y_2^-)}\,T(p_-,p_-';q) \,I_1\,I_2\,.}}
we have introduced integrals $I_{1,2}$,
\eqn\Io{\eqalign{I_1&=\int dp_+\, e^{ip_+\cdot (y_1^+-y_2^+)}\,P(p-q)P(p)\,,\cr
I_2&=\int dp_+'\, e^{-ip_+'\cdot (y_1^+-y_2^+)}\,P(p'-q)P(p')\,.}}
As discussed above, here we should set $y_1^+=y_2^+$.
%
In $\lambda_{IR}\rightarrow 0$ limit
\eqn\Iotr{I_1=I_2=\frac{i\pi^2\lambda_{IR}}{2\hat\lambda}\,,}
Using definition \xxpdef\ and equations \Tsolbs, \psinphin, \hatRdef, \Iotr, and neglecting overall factors of $\pi$, etc., we obtain
\eqn\Ko{\hat \KK_1(q)= \frac{1}{N}\int _0^1dx\,dx'\,e^{i(x-x')q_- (y_1^--y_2^-)}\,
\sum_r\frac{q_-^2}{q^2-\hat m_r^2}\,\phi_r(x)\phi_r^\star( x')\,.}
Now we need to plug \Ko\ into \Kyyo\ and integrate over $q$,
\eqn\Kott{\hat g^2\, \KK_1(y_1-y_2)= \hat\lambda\,\int _0^1dx\,dx'\int
d^2q\, \,e^{i(x-x')q_- (y_1^--y_2^-)}
\sum_r\frac{q_-^2}{q^2-\hat m_r^2},\phi_r(x)\phi_r^\star( x')}
We can change variables of integration (assuming some smooth regularization when $x=x'$,
which we discuss below), $q_-\rightarrow q_-/(x-x')$, $q_+\rightarrow (x-x')q_+$.
Then \Kott\ becomes
\eqn\Kotr{\hat g^2\, \KK_1(y_1-y_2)= \hat\lambda\int _0^1dx\,dx'\,\frac{1}{(x-x')^2}\,
 \int d^2q\,e^{iq_- (y_1^--y_2^-)}\sum_r\,\frac{q_-^2}{q^2-\hat m_r^2}\,\phi_r(x)\,\phi_r^\star( x')\,.}
Divergence in the integral in \Kotr\ when $x$ approaches $x'$ originates
from divergence in the IR limit, when momentum $p_--p_-'$, flowing between $y_1$ and $y_2$,
goes to zero. Therefore we can assume that such integral is regularized as in 't~Hooft equation,
by taking principal value part:
\eqn\intreg{a_r=-\int _0^1dx\,dx'\,\frac{1}{(x-x')^2}\,\phi_r(x)\,\phi_r^\star( x')=-\int _0^1dx'\,\phi_r^\star (x')\,\hat P\,
\int _0^1 dx\,\frac{\phi_r(x)}{(x-x')^2}\,.}
Using 't~Hooft equation \tHeq, normalization condition \tHeqsol\ of 't~Hooft wave-functions
$\phi_r(x)$, and definition \Mdrdef, we can re-write \intreg\ as
\eqn\intregt{a_r=\frac{\hat m_r^2}{\hat\lambda}+\left(1-\frac{\pi \hat m_q^2}{\hat\lambda}\right)\int _0^1dx\,|\phi_r(x)|^2
\frac{1}{x(1-x)}\,.}
The integral in \intregt\ is positive-valued, and for large $r$ we can use \philr\ and substitute
\eqn\intxomx{\int _0^1dx\,|\phi_r(x)|^2\,\frac{1}{x(1-x)}
\simeq \log (2\pi r)\,.}

In this subsection we have been considering the situation of generally non-vanishing
bare quark mass $m_q$. However, our computation of entanglement entropy
is performed in a set-up when we perturb CFT by four-fermionic interaction \SatHOOv.
If we had also perturbed CFT by non-vanishing bare mass for quarks, then an extra term,
besides \SatHOOv, would have been generated, and it would have also
modified the total contribution \pZZpntH\ to the entanglement entropy.
In particular, it is known that entanglement entropy of free massive fermion
is given by a non-trivial expression \refs{\CasiniFH,\CasiniSR}.
Therefore let us refrain to vanishing bare quark mass. Then from \intregt\
we obtain that $a_r>0$ for all $r$,
\eqn\intregtmz{a_r=\frac{\hat m_r^2}{\hat\lambda}+\int _0^1dx\,|\phi_r(x)|^2
\frac{1}{x(1-x)}\,.}

\subsec{Entanglement $c$-function}

We can use the result \Kotr\ in the expression \pZZttH,
\eqn\pZpNthr{\eqalign{\frac{1}{Z_1^H}\frac{\p\ZZ (n)}{\p n}\Bigg|_{n=1}=\hat \lambda\,\sum_ r\,a_r\int 
dy_{1,2}^-\,dy^+H_1(y_1,y_2)\int d^2q\,e^{iq_- (y_1^--y_2^-)}\frac{q_-^2}{q^2-\hat m_r^2}\,.}}
To find entanglement $c$-function in the 't~Hooft model we need to substitute \pZpNthr\
into \cdefon, and compute the integral. This is non-trivial, and requires regularization.
In fact, the integral which appears in \pZpNthr\ is exactly the same as the integral
which appears in analogous expression for the Schwinger model \pZZpntt.


To avoid the difficulty of calculation of integral in \pZpNthr, we use \ScU\ to obtain entanglement $c$-function in the 't~Hooft model, $c_H(\hat\lambda)$, in terms of entanglement
$c$-function of free massive scalars $c(\hat m_r)$
\eqn\ctH{c_H(\hat\lambda)=\frac{N}{3}+\sum_{r=2}^\infty\,a_r\,\frac{\hat\lambda}{\hat m_r^2}
\left(c\left(\hat m_r\right)-\frac{1}{3}\right)\,,}
where $\hat m_r=m_rL$. 
We know that $c_H(\hat\lambda)$ should be a monotonically decreasing function \CasiniCT,
which should fix unambiguously overall sign of the sum in \ctH\ to be plus
(at the end of previous Subsection it was shown that $a_r>0$, see \intregtmz).
Due to \intregtmz, \philr, \intxomx, we have $a_r\,\hat\lambda /\hat m_r^2=1+\OO(r^{-1}\,\log\,r)$ for large $r$.

The expression \ctH\ cannot be a final answer, and requires regularization. This can be consequence
of the way we handled integrals in \ScU\ and \pZpNthr.
Besides, as written now, \ctH\ has a contribution $N/3$ from twist operators, which diverges in the large-$N$
limit, and the sum over mesonic states, which is finite in the sense of large-$N$ counting.
To fix this problem, let us take non-trivial $L$-dependent part of \ctH, and
regularize the sum by requiring correct infra-red behavior.
Namely, we have $c_H|_{L=\infty}=1/3$,
which is the entanglement $c$-function for one free massless meson ($m_1=0$) in the spectrum.
This way we arrive at
\eqn\ctHIR{c_H(\lambda L^2)=\frac{1}{3}+ \sum_{r=2}^\infty\,a_r\,\frac{\hat\lambda}{\hat m_r^2}\,
c\left(\hat m_r\right)\,.}

Recall that \refs{\Casinihm,\CasiniSR} entanglement $c$-function of free massive scalar falls off exponentially
at large distances, $c(\hat m)\simeq e^{-\hat m}$. 
The masses of higher mesons are given by \philr. Therefore for interval of length $L$ the number of
mesons which contribute to the sum \ctHIR\ is
\eqn\mesm{\OO\left(\frac{1}{\lambda\,L^2}\right)\,.}
As $L$ is decreased, and we go to the ultra-violet, more mesons start to
contribute, and when $\lambda L^2=\OO(1/N)$ we get $c_H\simeq N$ due to \mesm,
with $\OO(N)$ light mesons contributing. In this regime of parametrically
small 't~Hooft coupling constant 
we can
neglect gauge interaction of quarks. Therefore we need to switch to picture of free
quarks, which gives $c_H=N/3$. It would be interesting to interpolate more smoothly
between the regimes of free quarks and mesons.

\newsec{Discussion}

In this paper we used the replica trick approach to
calculate entanglement $c$-function in three models of a two-dimensional
quantum field theory. We have considered Thirring model, Schwinger model, and
't~Hooft model. Thirring model is the simplest model with four-fermionic interaction,
and after bosonization it becomes re-formulated as a theory of free massless scalar.
Postponing bosonization, we have demonstrated explicitly that single-interval entanglement entropy
in the Thirring model is given by the CFT expression \CFTsiEE\ with central charge $c=1$.

We have shown that replica trick calculation of entanglement entropy in the Schwinger and 't~Hooft
models follows a similar path. In fact, one can derive entanglement entropy
in the 't~Hooft model, provided it is known what is the entanglement entropy
in the Schwinger model. The latter is indeed known (numerically
in general and analytically in asymptotic regions), because the Schwinger model is equivalent
to free massive scalar field.

It is essential for our calculation that 't~Hooft model is solvable in the large-$N$
limit. More precisely, it is important that one can compute quark-anti-quark scattering
amplitude, and demonstrate that the only intermediate states in this process
are mesons. Since each meson is free in the large-$N$ limit, we can use the known
result for entanglement entropy of free massive $2d$ scalar. 
We have demonstrated by explicit calculation that entanglement $c$-function in the 't~Hooft
model (in the large-$N$ limit) is equal to a sum of entanglement $c$-functions of all the mesons in the
spectrum (with certain numerical coefficients). It is interesting to observe that the fact
that each meson has a finite size, which moreover grows as the meson gets heavier, is not manifested
in the entanglement entropy. 

A possible explanation of the fact that each meson contributes to the entanglement entropy as
a point-like particle is the following. As it was mentioned
in Introduction, defining entanglement
entropy in systems with gauge fields is subtle.
The problem appears with finite-size gauge-invariant
operators (like mesons in the 't~Hooft model) which are partly located in $\AA$ and partly in $\bar\AA$.
Such operators are in conflict with the very definition of entanglement entropy, which
starts with the assumption of splitting of the Hilbert space into a tensor product
\eqn\Hspaspl{\HH_\AA\times \HH_{\bar\AA}\,.}
Therefore the question is what exactly the replica trick calculation, which we performed in this paper,
has given us.

One answer is that in the entanglement entropy calculation the gauge-invariant
operators, which are located partly in $\AA$ and partly in $\bar\AA$, are cut
into two parts, by the entangling surface. The free ends of those parts are glued to the extra added
gauge non-invariant (charged) edge modes, located
on the entangling surface \refs{\DonnellyGVA,\DonnellyFUA} (see also \HuangPFA\
for the calculation of the contribution of the edge modes to the entanglement entropy).
Therefore each of the resulting parts is completed to a gauge-invariant
object. One is located entirely in $\AA$, the other is located entirely in $\bar\AA$.
This procedure is incorporated in the replica trick calculation \DonnellyGVA.
This means that the whole Hilbert space is enhanced
to include such charged boundary degrees of freedom,
located on the entangling surface. The consequence of such an enhancement is that
there are no more gauge-invariant operators, located partly in $\AA$ and partly in $\bar\AA$.
This makes definition of the entanglement entropy consistent. At the same time it explains why
finite size of mesons is not manifested in the entanglement $c$-function, no matter how long the meson is.
\foot{In a pure two-dimensional Yang-Mills theory there is an additive contribution to the
entanglement entropy, which can be interpreted by counting the gauge
non-invariant states, located on the entangling surface \DonnellyGVA. See also \KabatJQ,
where entanglement entropy in the $O(N)$ $\sigma$-model has been calculated,
and interpretation in terms of counting the states of the UV degrees of freedom
has been provided.
It might be that there are such parton-counting terms, entering entanglement entropy in the 't~Hooft model,
and these terms are $L$-independent, and hence invisible in the calculation of the entanglement $c$-function.
We thank A.~Wall for discussion of this point.}

If bare quark mass is zero, then the lowest meson
is massless. Therefore in the deep infra-red we have CFT of one free massless scalar,
and entanglement $c$-function is equal to $1/3$. 
In the deep ultra-violet regime quarks are free, and
entanglement $c$-function is equal to $N/3$, where $N$ is the number of colors.
One can compare this crossover from the mesonic phase to free quark phase
with the phase transition, observed in \KlebanovWS\ (see also \LewkowyczMW),
where single-interval (slab) entanglement entropy in confining gauge theory has been calculated holographically.
In \KlebanovWS\ it was shown that there is a critical thickness of the slab $L_c$ so that when $L<L_c$ entanglement entropy behaves as $N^2$,
and when $L>L_c$ entanglement entropy behaves as $N^0$. It would be interesting to understand
better how these two observations are related.

It would be interesting to address the question, raised in the introduction, and
calculate entanglement entropy in the quark language, perturbatively in the coupling constant.
Such calculation will shed light on the crossover region, $\lambda L^2=1/N$.

\bigskip
\bigskip
{\noindent \bf Acknowledgements:}
I would like to thank B.~Galilo, C.~Herzog, K.-W.~Huang, D.~Kutasov, S.~Lee,
E.-G.~Moon, A.~Parnachev, N.~Poovuttikul, M.~Roberts, M.~Smolkin, D.~Son and A.~Wall for useful discussions and correspondence. I would like to thank D.~Kutasov for suggesting the problem,
discussions and collaboration on every stage of this project.
I also would like to thank D.~Kutasov and A.~Parnachev for reading and commenting the draft of this paper.
This work was supported by Oehme Fellowship.

\appendix{A}{Gap equation for fermionic propagator}

In this Appendix we derive gap equation on full quark propagator by calculating path integral in the large-$N$
limit. This is different from the conventional derivation, which involves
equation satisfied by planar ladder diagrams \tHooftHX.
Similar derivation for Gross-Neveu model can be found in \refs{\AntonyanQY,\AntonyanVW}.

For generality, let us give fermions the bare mass $m$. The total action is then
\eqn\SthC{S=S_0+S_{int}\,,}
where
\eqn\SntH{S_0=\int d^2z\,\hat\Psi^{ \alpha\dagger} (z)(\gamma^\mu\p_\mu +m)\hat\Psi^\alpha (z)}
and (let us absorb 't~Hooft coupling $\lambda$ into rescaling of coordinates $z_{1,2}$ and the mass $m$)
\eqn\SitH{S_{int}=\frac{1}{N}\int d^2z_{1,2}\,G(z_1-z_2)\Psi^{\alpha\star}(z_1)\Psi^\alpha (z_2)
\Psi^{\beta\star}(z_2)\Psi^\beta (z_1)\,.}
We have denoted Dirac fermion as $\hat\Psi$, and its {\it charge} conjugate as $\hat\Psi^\dagger$,
\eqn\PsiPsi{\hat\Psi=\left({\Psi\atop \bar\Psi}\right)\,,\quad\quad \hat\Psi^\dagger=\left(\bar\Psi^\star\,,\;\Psi^\star\right)\,,}
and the Dirac operator is
\eqn\gammad{\gamma^\mu\p_\mu=\left({0\atop\bar\p}\;{\p\atop 0}\right)\,.}
Let us introduce auxiliary gauge-singlet field $P(z_1,z_2)$, which for the purpose of finding
quark two-point function, due to translational symmetry, is taken to be $P(z_1-z_2)$. We can re-write the action \SitH\
as
\eqn\SiP{S_{int}=\int d^2z_{1,2}\,G(z_1-z_2)\left(NP(z_1-z_2)P(z_2-z_1)
+P(z_1-z_2)\Psi^{\alpha\star}(z_1)\Psi^\alpha (z_2)\right)}
Then the total action is the sum of quadratic action for $P(z_1-z_2)$ and the fermionic action:
\eqn\StotPP{\eqalign{
S&=S_\Psi+S_P\,,\cr
S_\Psi&=\int d^2z_{1,2}\,\hat\Psi^{\alpha\dagger} (z_1)
\left[\delta(z_1-z_2)\gamma^\mu\p_\mu+\delta(z_1-z_2)\,m+\left({0\atop 1}\;{0\atop 0}\right)
P(z_1-z_2)G(z_1-z_2)\right]
\hat\Psi^\alpha(z_2)\,,\cr
S_P&=N\int d^2z_{1,2}\,G(z_1-z_2)P(z_1-z_2)P(z_2-z_1)\,.
}}
In momentum representation
\eqn\Smomrep{\eqalign{
S_\Psi&=\int d^2q\,\hat\Psi^{\alpha\dagger} (-q,-\bar q)\,K\,\hat\Psi^\alpha (q,\bar q)\,,\cr
K&=\left({m\atop{q+T(q)}}\;\;{\bar q\atop m}\right)\,,\qquad T(q)=\int d^2k\frac{P(k,\bar k)}{(q-k)^2}\,,\cr
S_P&=N\,\int d^2z\int d^2q\,d^2k\,\frac{P(k,\bar k)P(q,\bar q)}{(q-k)^2}\,.
}}
Integrating over fermions gives
\eqn\Zpsiths{Z_\Psi=\int [d\Psi]e^{-S_\Psi}=e^{N\,\Tr\,\log\, K}\,,}
where
\eqn\TrlogK{\Tr\,\log\, K=\int d^2z\,d^2q\,\log\,\det\,K(q,\bar q)=
\int d^2z\,d^2q\,\log\,\left(m^2-|q|^2-\bar q\int d^2k\frac{P(k,\bar k)}{(q-k)^2}\right)}
Combining \TrlogK\ with $S_P$ from \Smomrep\ we obtain the total action for $P(q,\bar q)$:
\eqn\SPtot{S=N\,\int d^2zd^2q\,\left[\int d^2k\frac{P(k,\bar k)P(q,\bar q)}{(q-k)^2}
-\log \,\left(m^2-|q|^2-\bar q\int d^2k\,\frac{P(k,\bar k)}{(q-k)^2}\right)\right]\,.}
In the large-$N$ limit solution to the theory is obtained simply by extremization of \SPtot\
w.r.t. $P(q)$:
\eqn\DP{\delta S=N\,\int d^2zd^2qd^2k_1\frac{\delta P(k_1,\bar k_1)}{(q-k_1)^2}
\left[P(q,\bar q)-\frac{\bar q}{|q|^2-m^2-\bar q\int d^2k\,\frac{P(k,\bar k)}{(q-k)^2}}\right]}
The corresponding equation of motion is
\eqn\Peom{P(q,\bar q)=\frac{\bar q}{|q|^2-m^2-\bar q\int d^2k\,\frac{P(k,\bar k)}{(q-k)^2}}\,.}

We can see that the equation \Peom\ is just 't~Hooft gap equation \tHooftHX\ on full fermionic
propagator $P(q,\bar q)$ in leading order in large-$N$ limit.
Indeed, the full fermionic propagator is equal to the sum of geometric series of 1PI
propagators. On the other hand, 1PI propagator is equal to \tHooftHX\
\eqn\oPI{\Sigma (q,\bar q)=\int d^2k\,\frac{P(k,\bar k)}{(q-k)^2}\,}
and therefore the sum of geometric series is equal to
\eqn\sumgs{\frac{\frac{\bar q}{|q|^2-m^2}}{1-\frac{\bar q}{|q|^2-m^2}\Sigma(q,\bar q)}=
\frac{\bar q}{|q|^2-m^2-\bar q\int d^2k\,\frac{P(k,\bar k)}{(q-k)^2}}\,.}
This agrees with the equation \Peom.

\appendix{B}{Bethe-Salpeter equation for four-quark vertex}

Let us give a short review of how one can arrive at the Bethe-Saltpeter equation \BSTeq\
on four-quark vertex $T$. Consider four quarks scattering process,
with two incoming quarks
\eqn\Binq{\left(\hat\psi^{\alpha _1\star}(p_1)\,,\; \hat\psi^{\alpha_1}(p_2)\right)}
where $\alpha_1$ is a fixed color, and two outgoing quarks
\eqn\outgqB{\left(\hat\psi^{\alpha_3}(p_3)\,,\; \hat\psi^{\alpha_3\star}(p_4)\right)\,.}
Denote $q=p_1-p_2$ to be momentum flowing in this process.
Due to \drverdef\ such diagram is equal to
\eqn\Bo{\delta(p_1-p_2+p_3-p_4)P(p_1)P(p_2)P(p_3)P(p_4)T(p_1,p_3;q)\,.}

On the other hand \CallanPS, this diagram is equal to sum of the diagram
\eqn\Bt{\delta(p_1-p_2+p_3-p_4)P(p_1)P(p_2)P(p_3)P(p_4)\frac{ig^2}{(p_1-p_3)^2}\,,}
describing scattering process in which incoming quarks \Binq\ exchange one gluon and fly away as \outgqB;
and the diagram
\eqn\Btt{\delta(p_1-p_2+p_3-p_4)\,P(p_1)P(p_2)\,\sum_{\alpha_1'}\int d^2k\,i g^2\,P(k)P(k-q)\,\frac{1}{(k-p_1)^2}\,
T(k,p_3;q)
P(p_3)P(p_4)\,,}
describing process in which two incoming quarks \Binq\ exchange one gluon with momentum $k-p_1$,
scatter in two outgoing quarks
\eqn\Binq{\left(\hat\psi^{\alpha _1'\star}(k-q)\,,\; \hat\psi^{\alpha_1'}(k)\right)\,,}
which subsequently scatter on vertex $T(k,p_3;q)$, giving two outgoing quarks \outgqB.
This way we obtain equation \BSTeq.

\footatend\vfill\supereject\immediate\closeout\rfile\writestoppt
\baselineskip=14pt\centerline{{\bf References}}\bigskip{\frenchspacing%
\parindent=20pt\escapechar=` \input refs.tmp\vfill\eject}\nonfrenchspacing
\end